\documentclass[preprint]{iucr} % DO NOT DELETE THIS LINE
\usepackage[utf8]{inputenc}
\usepackage{graphicx}
\usepackage{textcomp}
\usepackage{palatino}
\usepackage{gensymb}
\usepackage{siunitx}
\newcommand{\angstrom}{\textup{\AA}}
\usepackage{multirow}
\usepackage{tabularx}

     \journalcode{S}              % Indicate the journal to which submitted
\begin{document}                  % DO NOT DELETE THIS LINE

     %-------------------------------------------------------------------------
     % The introductory (header) part of the paper
     %-------------------------------------------------------------------------

     % The title of the paper. Use \shorttitle to indicate an abbreviated title
     % for use in running heads (you will need to uncomment it).

\title{I22: SAXS/WAXS beamline at Diamond Light Source - an overview of 10 years operation}
%\shorttitle{Short Title}

     % Authors' names and addresses. Use \cauthor for the main (contact) author.
     % Use \author for all other authors. Use \aff for authors' affiliations.
     % Use lower-case letters in square brackets to link authors to their
     % affiliations; if there is only one affiliation address, remove the [a].

\author[a]{A. J.}{Smith}
\author[a]{S. G.}{Alcock}
\author[a]{L. S.}{Davidson}
\author[a]{J. H.}{Emmins}
\author[d]{J. C.}{Hiller Bardsley}
\author[a]{P.}{Holloway}
\author[c]{M.}{Malfois}
\author[a]{A. R.}{Marshall}
\author[a]{C. L.}{Pizzey}
\author[b]{S. E.}{Rogers}
\author[a]{O.}{Shebanova}
\author[a]{T.}{Snow}
\author[a]{J. P.}{Sutter}
\author[a]{E. P.}{Williams}
\cauthor[a]{N. J.}{Terrill}{nick.terrill@diamond.ac.uk}{address if different from\aff}

\aff[a]{Diamond Light Source Ltd, Diamond House, Harwell Science and Innovation Campus, Didcot, Oxfordshire, OX11 0DE, \country{United Kingdom}}
\aff[b]{ISIS Neutron and Muon Source, Science and Technology Facilities Council, Rutherford Appleton Laboratory, Didcot, Oxfordshire, OX11 0QX, \country{United Kingdom}}
\aff[c]{ALBA Synchrotron, Carrer de la Llum 2-26, 08290 Cerdanyola del Vall\`{e}s, Barcelona, \country{Spain}}
\aff[d]{King's College London, Guy's Campus, Great Maze Pond, London SE1 1UL \country{United Kingdom}}

     % Use \shortauthor to indicate an abbreviated author list for use in
     % running heads (you will need to uncomment it).

%\shortauthor{Soape, Author and Doe}

     % Use \vita if required to give biographical details (for authors of
     % invited review papers only). Uncomment it.

%\vita{Author's biography}

     % Keywords (required for Journal of Synchrotron Radiation only)
     % Use the \keyword macro for each word or phrase, e.g. 
     % \keyword{X-ray diffraction}\keyword{muscle}

%\keyword{keyword}

     % PDB and NDB reference codes for structures referenced in the article and
     % deposited with the Protein Data Bank and Nucleic Acids Database (Acta
     % Crystallographica Section D). Repeat for each separate structure e.g
     % \PDBref[dethiobiotin synthetase]{1byi} \NDBref[d(G$_4$CGC$_4$)]{ad0002}

%\PDBref[optional name]{refcode}
%\NDBref[optional name]{refcode}

\maketitle                        % DO NOT DELETE THIS LINE

\begin{synopsis}
Beamline I22, a versatile SAXS/WAXS beamline at Diamond Light Source is presented, along with an overview of 10 years operation of the beamline.
\end{synopsis}

\begin{abstract}
Beamline I22 at Diamond Light Source is dedicated to the study of soft matter systems from both biological and materials science. The beamline can operate in the range \SIrange{3.7}{22}{\kilo\electronvolt} for transmission SAXS and \SIrange{14}{20}{\kilo\electronvolt} for microfocus SAXS with beamsizes 240 x 60 \si{\micro\metre\squared} spot [Full width half maximum (FWHM) Horizontal (H) x Vertical (V)] at sample for the main beamline, and approximately 10 x 10 \si{\micro\metre\squared} for the dedicated microfocussing platform. There is a versatile sample platform for accommodating a range of facilities, and user developed, sample environments. The high brilliance of the insertion device source on I22 allows structural investigation of materials under extreme environments (for example, fluid flow at high pressures and temperatures). I22 provides reliable access to millisecond data acquisition timescales, essential to understanding kinetic processes such as early folding events in proteins or structural evolution in polymers and colloids.
\end{abstract}

     %-------------------------------------------------------------------------
     % The main body of the paper
     %-------------------------------------------------------------------------
     % Now enter the text of the document in multiple \section's, \subsection's
     % and \subsubsection's as required.

\section{Introduction}

Small Angle X-ray Scattering (SAXS) provides essential information on the structure and dynamics of large molecular assemblies in low order environments. These are characteristic of living organisms and also many complex materials such as polymers and colloids. Relevant active research in the UK encompasses the fields of medicine \cite{Ma2016,burton2019,coudrillier2016,Al-Jaibaji2018,kudsiova2019}, biology \cite{Troilo2016,McGeehan2011,Arnold2011,Salamah2018}, the environment \cite{Neill2018,Seddon2016} and materials \cite{Summerton2018a,Wychowaniec2018,Burton2017}, and includes studies of supramolecular organization in biomechanical systems \cite{Xi2018,Kampourakis2018,Sui2014a}, corneal transparency \cite{Morgan2018,Hayes2017}, biological membranes \cite{Barriga2016,Slatter2018,Tang2014}, polymer processing \cite{Stasiak2015,Wan2018,Heeley2013,Toolan2017}, colloids \cite{Calabrese2018,Poulos2016,Mable2016}, inorganic aggregates \cite{Raine2018,Bennett2015,Zhou2018}, liquid crystals \cite{Hallett2014,Prehm2018,Lehman2018}, and devices \cite{Xia2017, Barrows2016}.

\section{Beamline overview}

The scientific and technological challenges confronted by this diverse community required a high resolution, high brightness, synchrotron beamline. The first small angle scattering beamline at Diamond, I22, uses an in-vacuum undulator source to deliver a high photon flux into a focused 240 x 60 \si{\micro\metre\squared} spot [Full width half maximum (FWHM) Horizontal (H) x Vertical (V)] at sample for the main beamline, and approximately 10 x 10 \si{\micro\metre\squared} for a dedicated microfocussing platform. This platform will the subject of a further technical paper, and will not be discussed in detail here. The main beamline has the potential for a continuous energy range of \SIrange{3.7}{22}{\kilo\electronvolt} with, at present, a reduced range of \SIrange{14}{20}{\kilo\electronvolt} for the microfocus option. It is currently operated from \SIrange{7}{20}{\kilo\electronvolt}; lower energy operation would require either helium or vacuum sample chambers. Depending on the energy used, and the exact geometry of the scattering experiment, the q range achievable at I22 is $0.0011 \leq q(\angstrom^{-1}) \leq 9.45$. The SAXS camera is composed of evacuated sections (\SI{5e-5}{\milli\bar} of flight tube with a nosecone incorporating the WAXS detector at the sample end and ended by a 310mm diameter Kapton Window at the SAXS detector end). These tubes can be connected, as required, to form camera lengths anywhere from 1.9 to 9.9m in 0.25m steps. The primary end station, with associated area detectors for static and time resolved measurements, is capable of recording the scattered radiation from samples contained in a range of commercial and bespoke sample environments.

\begin{figure}
	\begin{center}
		\includegraphics[width=0.9\textwidth]{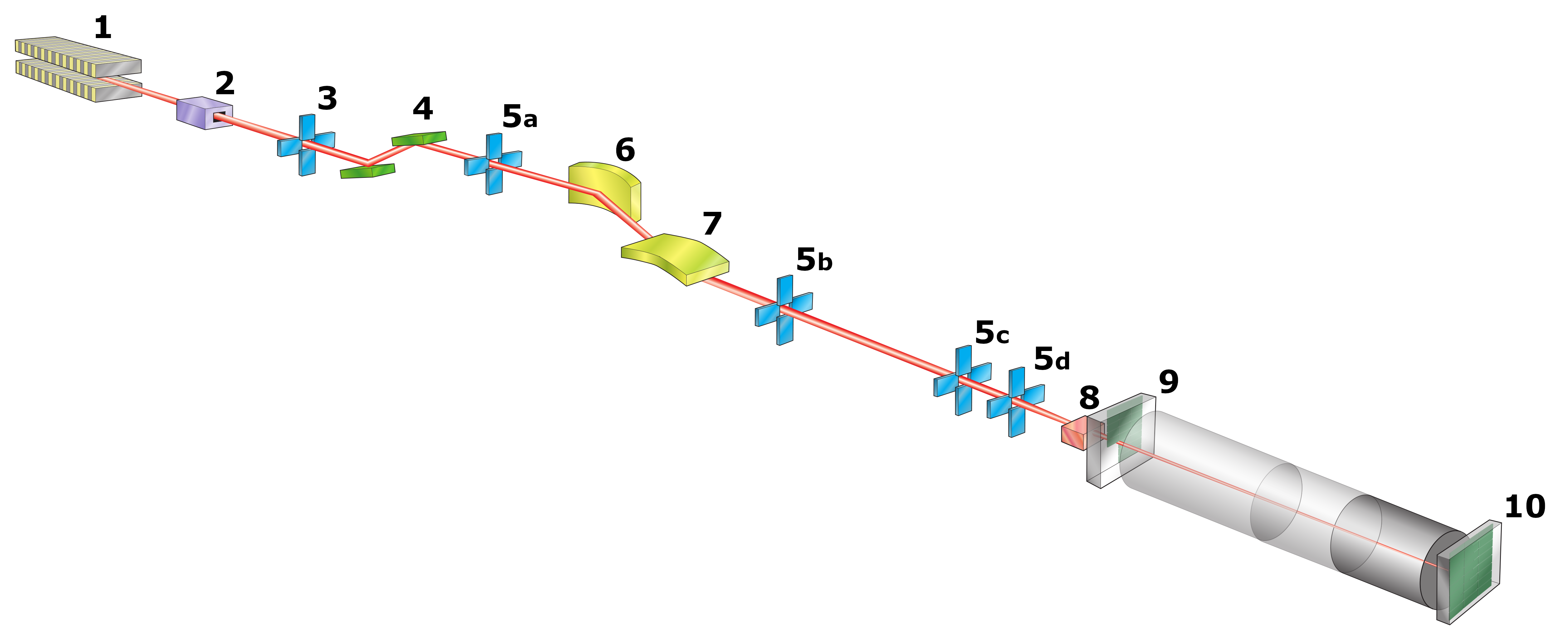}
		\caption{Layout of I22}\label{fg:BL_layout}
	\end{center}
\end{figure}

\begin{table}
    \centering
    \begin{tabularx}{400pt}{cll}      % Alignment for each cell: l=left, c=center, r=right
        Figure 1 index    & Component & Distance from source (m) \\
        \hline
        1 & in-vacuum undulator 25mm period & 0.00 \\
        2 & customised aperture  & 17.20\\
        3 & white beam slits & 22.70 \\
        4 & double crystal monochromator (Si 111) & 24.00 \\
        5 a-d & monochromatic slits & 25.57, 28.55, 46.00, 47.00  \\
        6 & horizontal focussing mirror & 26.78\\
        7 & vertical focussing mirror & 27.73\\
        8 & sample position & \SIrange{47.1}{48}{}\\
        9 & wide angle detector & 0.17 (from sample)\\
        10 & small angle detector & \SIrange{1.9}{9.9}{} (from sample)\\[1em]
    \end{tabularx}
    
    \caption{Major components of I22\\[1em]}
    \label{tb:components}
\end{table}

\section{Undulator}

\begin{figure}
	\begin{center}
		\includegraphics[width=0.9\textwidth]{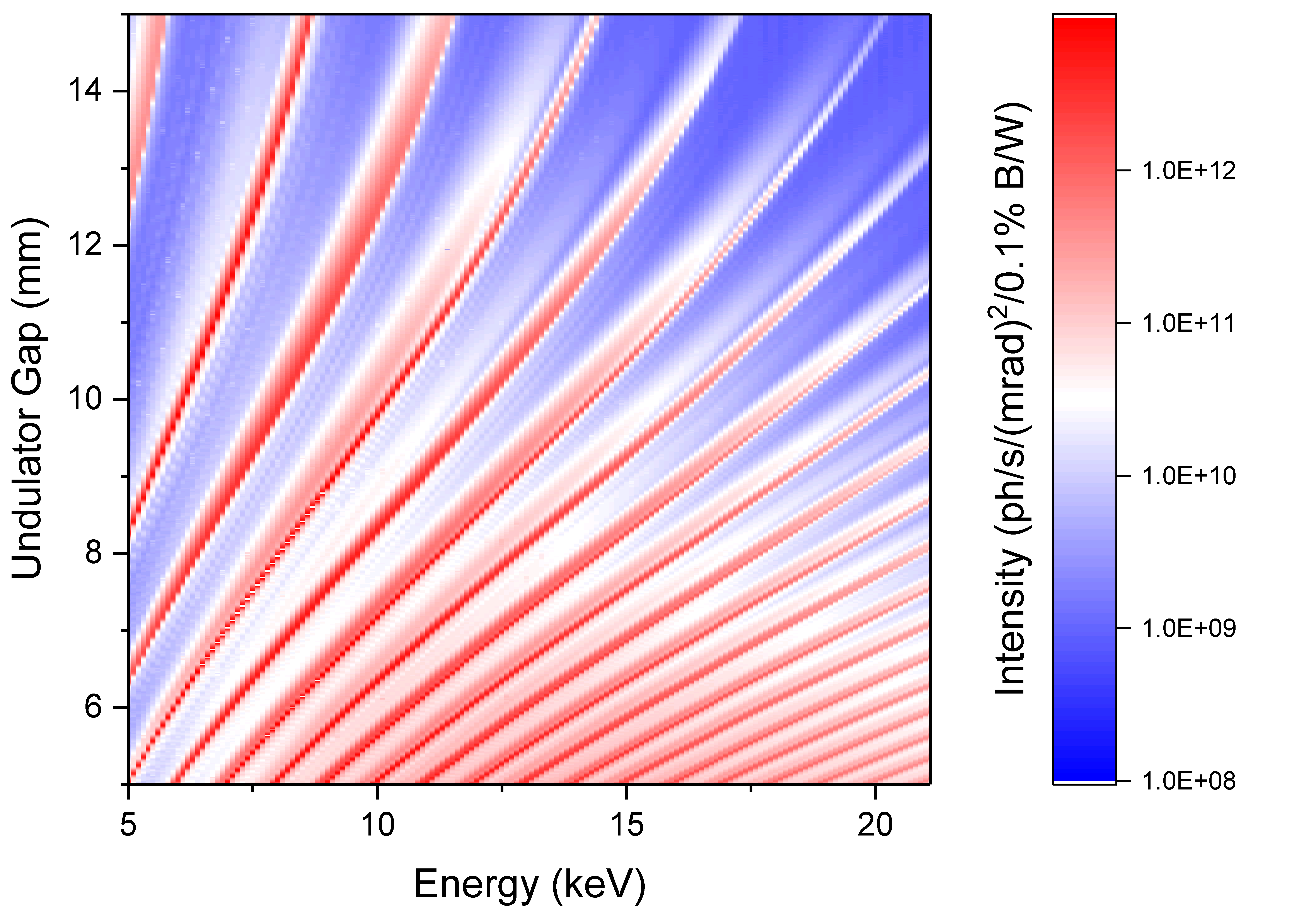}
		\caption{In-vacuum U25 undulator spectrum of I22. Flux values measured with calibrated diagnostic after the monochromator. N.B. The primary White Beam slits were closed to a central 100 x 100 \si{\micro\metre\squared} gap to give clean undulator spectrum }\label{fg:undulator_spectrum}
	\end{center}
\end{figure}

The beamline operates a, Diamond designed \cite{Patel2017}, 2m in-vacuum undulator with a period of 25 mm, providing continuous energy coverage over the energy range of the beamline. The beamline is optimised for the energy range \SIrange{8}{20}{\kilo\electronvolt} working primarily with the 5$^\text{th}$ through to the 17$^\text{th}$ harmonics. Off-axis undulator radiation is removed to minimise heat load on the beamline optics, via a water-cooled 150 x 75 \si{\micro\radian} (H x V) aperture located inside the storage ring tunnel at \SI{17.202}{\metre} downstream from the source. Water-cooled primary slits further define the beam before energy selection and focusing. The undulator has a very small phase error (\SI{0.02}{\milli\metre}). Figure \ref{fg:undulator_spectrum} shows the spectrum as a function of energy and undulator gap. All available harmonics can clearly be distinguished, from the 2$^\text{nd}$ at the top left, through to he 20$^\text{th}$ at the bottom right.  

\section{Monochromator}
The monochromator (Oxford Danfysik, Oxford, UK) is a fixed-exit double-crystal design, originally established for 3 crystal sets. It is currently installed with a Si(111) crystal pair; this allows access over the full operational energy range, with a resolution of 1.6 x $10^{-4}$ $\Delta E/E$ at \SI{10.0}{\kilo\electronvolt}. Vertical translation of the second crystal maintains a fixed beam exit configuration for the beamline. Both the first and second crystals are indirectly cooled by liquid nitrogen, to prevent damage arising from the high power density of the undulator. An indium foil provides thermal contact for both crystals to a cooled copper block. The first crystal is cooled by liquid nitrogen flowing continuously through the copper block while heat from the second crystal is removed via copper braiding. The Diamond storage ring has run in top up mode since October 2008, with an injection every ten minutes. This produces a very slight variation in ring current of $\sim$2\% or $\pm$\SI{5}{\milli\ampere} at most, therefore changes in the heat load on the monochromator during an experiment are primarily due to undulator gap and harmonic changes associated with a change in energy. Both coarse motors ($\pm$\SI{4}{\milli\radian}) and fine piezo actuators ($\pm$\SI{180}{\micro\radian}) are used to adjust the pitch and roll of the crystals, maintaining alignment of the first and second crystal lattice planes.  

\subsection{Monochromator calibration}
\begin{figure}
	\begin{center}
	    \includegraphics[width=0.9\textwidth]{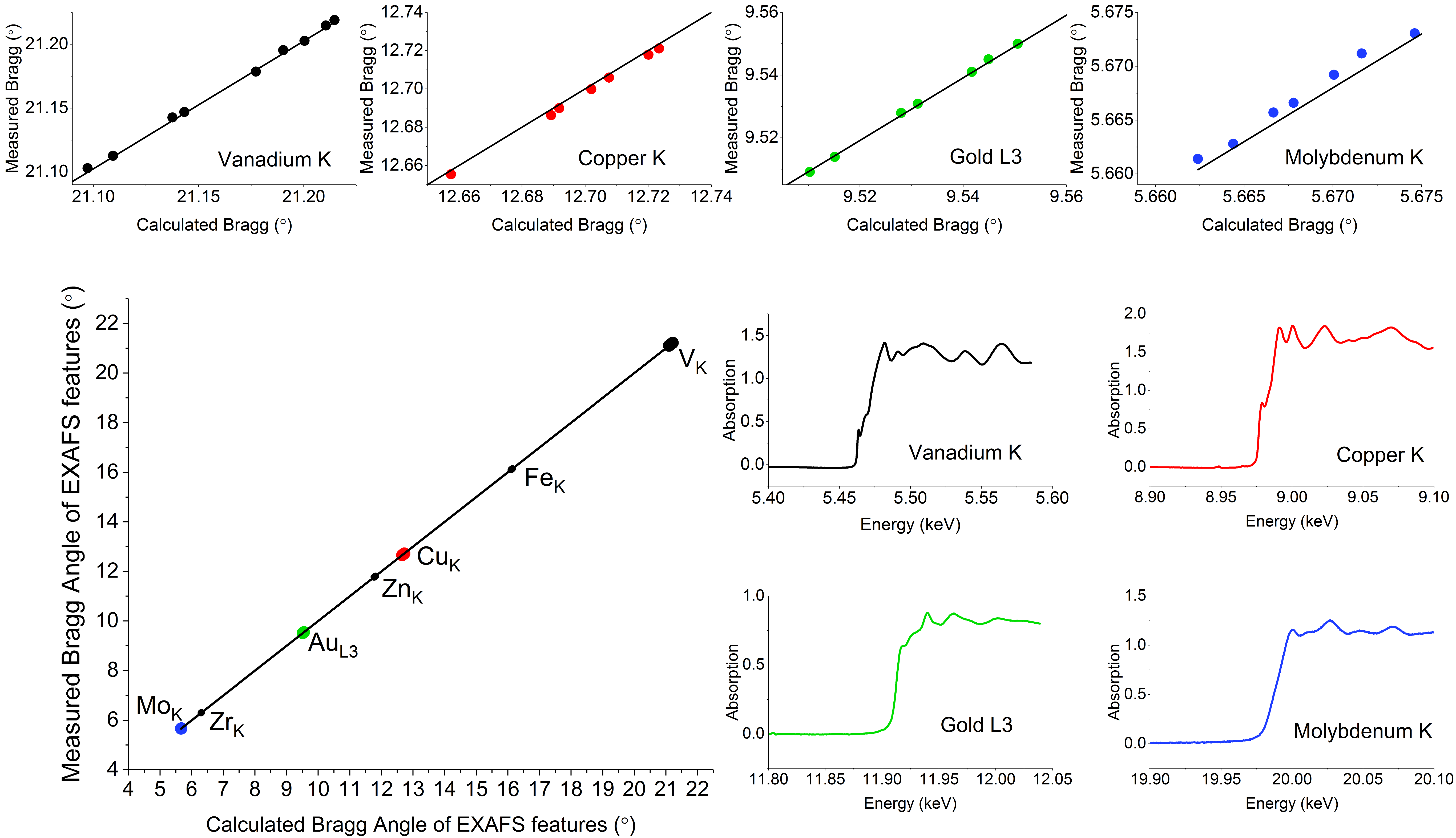}
		\caption{Calibration of the I22 monochromator using EXAFS from metal foils}\label{fg:MonoEXAFS}
	\end{center}
\end{figure}

The monochromator is periodically calibrated using EXAFS spectra collected from a series of retractable metal foils that are permanently installed in the beamline downstream of the monochromator. Scatter diodes before and after the foils record $I_0$ and $I_t$ values, respectively. EXAFS features from the spectra (peaks), and the derivative of the spectra (edges) for all of the foils (V, Fe, Cu, Zn, Au, Zr, and Mo) are fitted, against their expected positions from literature values, \cite{BEARDEN1967}
in a global least squares minimisation. The main part, lower left, of Figure \ref{fg:MonoEXAFS} shows the overall fit. The lower right section of Figure \ref{fg:MonoEXAFS} shows the absorption spectra from V, Cu, Au, and Mo as measured on I22. Expanded views of the linear fitting of the vanadium, copper, gold, and molybdenum features are shown along the top, demonstrating the consistency of the fit across a wide range of Bragg angles. In a typical fit 46 data points, spanning \ang{15.55}, are used to calibrate the Bragg axis.

%fix these values

As a way of evaluating the repeatability 10 repeat spectra were measured for all of the above edges. The data is given in the supplementary information. Standard deviations for the measured positions of EXAFS features of  \SI{3.8}{\micro\radian} at the vanadium edge, and \SI{1.8}{\micro\radian} at the molybdenum edge were found highlighting the excellent repeatability of the monochromator design.

\section{Mirrors}

The beamline focusing comprises a Kirkpatrick–Baez (KB) \cite{Kirkpatrick1948} mirror pair (ACCEL, Germany), using adaptive bimorph mirror technology, \cite{Susini1995,Cautero2007} to provide independent horizontal and vertical focusing at any point in the end station while still operating close to 1:1 focusing. Bimorph mirrors have three advantages that make them useful for I22. First, by varying the voltages on individual electrodes, one can correct the waviness left on the mirror’s surface by the polishing process, thus achieving a sharp focal spot with small tails. Second, a bimorph responds within seconds to changes in voltage, and even settling times imposed by imperfect mounting of the bimorph generally do not exceed 10-15 minutes. Therefore, bimorph mirrors permit the size, shape and focal distance of the X-ray beam to be changed rapidly to accommodate different camera lengths. Third, they have good long-term stability, and thus can be left at a constant voltage setting over weeks or months without loss of focal quality. The horizontal focusing mirror (\SI{900}{\milli\metre} long, with an \SI{820}{\milli\metre} active length) and the vertical focusing mirror (\SI{600}{\milli\metre} long, with a \SI{550}{\milli\metre} active length) are located at \SI{26.8}{\metre} and \SI{27.7}{\metre} respectively from the source and share a common vacuum vessel. Each mirror substrate is made of fused silica, \SI{50}{\milli\metre} wide with a central \SI{15}{\milli\metre} wide coated stripe of rhodium; the stripes provide excellent harmonic rejection across the beamline's energy range while operated at \SI{2.6}{\milli\radian}. The adaptive bimorph capability is provided by piezo actuators attached to each substrate. The vertical mirror has 32 electrodes, which are paired along the length to give 16 virtual electrodes. These are used to correct the mirror figure and provide vertical focusing. The horizontal mirror has 12 electrodes, 4 at either end are coupled into 2 pairs and the 4 remaining electrodes are left as singles to give finer control over the central portion of the mirror, where it is of greatest benefit for providing optimal horizontal focusing. 

\subsection{Mirror refurbishment}

\begin{figure}
	\begin{center}
		\includegraphics[width=0.9\textwidth]{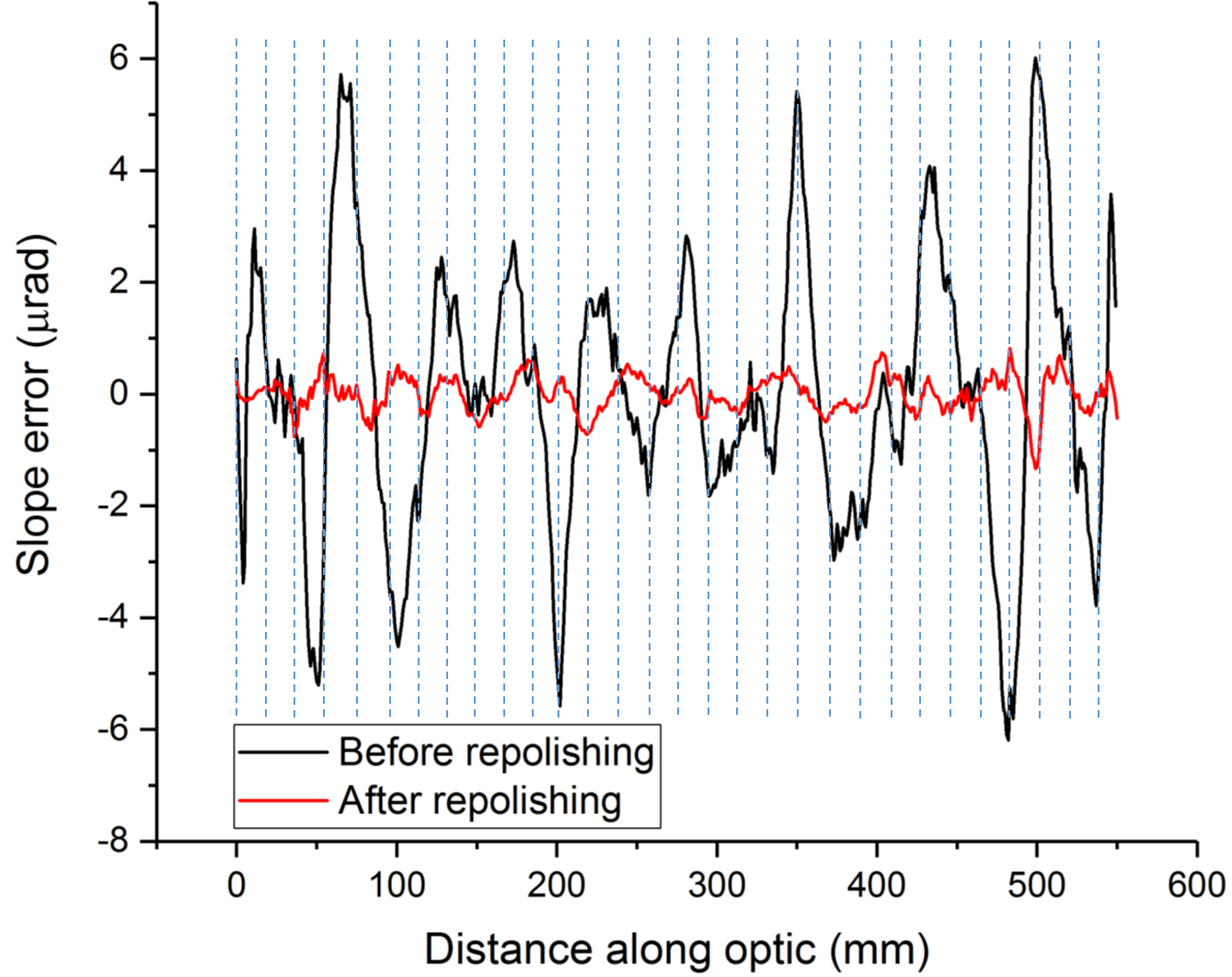}
		\caption{Effect of repolishing the I22 VFM as determined by Diamond NOM}\label{fg:repolish}
	\end{center}
\end{figure}

Examination of the Bimorph mirrors using the x-ray beam and the Diamond Light Source Nanometre Optical Metrology (NOM) \cite{Alcock2010} non-contact profiler revealed damage at the piezo interfaces \cite{Alcock2013} as indicated by the gross changes in slope error across the mirror in the measurements shown in Figure \ref{fg:repolish}. While not all junctions behaved the same (dotted lines in \ref{fg:repolish} show the position of the junctions, and features are invariably found near to the junctions), these effects translated into poor focusing performance rendering optimum focusing impossible and a degradation in beamline background. 
Repolishing the optical surface of both mirrors significantly improved the focusing capability of the mirrors to the point where optimum focus is now achievable. Data from the Diamond NOM shows a 9-fold improvement in slope error after repolishing from \SIrange{2.7}{0.3}{\micro\radian}. 

We stress that the junction effect has not been found to appear in the latest “second-generation” bimorph mirrors, which have their piezoelectric plates attached to the sides of the mirror substrate rather than being sandwiched between an upper and a lower substrate, as is the case with I22’s first-generation bimorph mirrors\cite{Alcock2019,Alcock2019a}.

The table below gives an indication of the beamsize at various camera lengths on I22 after correcting the mirror surfaces.

% TS - To look into why this isn't centering
\begin{table}
    \centering
    \begin{tabularx}{400pt}{>{\raggedright\arraybackslash}X>{\centering\arraybackslash}X>{\centering\arraybackslash}X>{\centering\arraybackslash}X>{\centering\arraybackslash}X>{\centering\arraybackslash}X>{\centering}X}
        Focal Position    & Demagnified FWHM source size (HxV, \si{\micro\metre})  & H FWHM slope error broadening (\si{\micro\metre}) & V FWHM slope error broadening (\si{\micro\metre}) & Measured FWHM focal size (HxV, \si{\micro\metre})\\
        \hline
        Sample & 258 x 6 & 39 & 60 & 262 x 60 \\
        Sample + 3m	& 296 x 7	& 45  & 69 & 299 x 69 \\
        Sample + 5m	& 321 x 7	& 49 & 75 & 329 x 75\\
        Sample + 7m	& 346 x 8	& 52 & 81 & 350 x 81 \\[1em]
    \end{tabularx}

    \caption{Beamsizes for the repolished Bimorph Mirrors\\[1em]}
    \label{tb:beamsize}
\end{table}

These measurements do not perfectly match the theoretical widths. However, with reasonable values of electron beam size and mirror slope error, the theoretical widths come within 16\% of the measured horizontal widths, and 10\% of the measured vertical widths (Table \ref{tb:beamsize}).

\section{Endstation}

\subsection{Flux calibration}

Flux values are reported below for a selection of commonly used experimental conditions. The measurements were taken using with a calibrated diode (Canbera PD-300 PIPS diode calibrated by PTB Berlin) at the sample position.

\begin{table}
    \centering
    \begin{tabular}{cc}
        Energy    & Measured Flux \\
        / \si{keV} & photons \si{\per\second} \\
        \hline
        7 & \num{1.08e13} \\
        10 & \num{6.94e12} \\
        12.4	& \num{3.97e12} \\
        14	&  \num{2.74e12}\\
        18	& \num{1.48e12} \\[1em]
    \end{tabular}
    
    \caption{Measured Flux for commonly used energies\\[1em]}
    \label{tb:flux}
\end{table}

\subsection{Sample environments}

Samples are mounted on a platform (IDT, Cheshire, UK; customised design) allowing remotely controlled motion of \SI{200}{\milli\metre} both horizontally and vertically. The platform is mounted on rails for manually controlled motion of \SI{1000}{\milli\metre} along the beam path. Combined with motorised X,Y and manual Z motion of the WAXS detector nosecone, any reasonable size of sample environment can be accommodated whilst minimising air gaps and attendant background scatter. The sample table surface comprises a standard optical breadboard (750 x 750 \si{\milli\metre\squared}, M6 tapped holes on a \SI{25}{\milli\metre} pitch) which facilitates mounting of sample environments, large and small. The platform has \SI{5}{\micro\metre} resolution, and can accommodate large sample environments (up to \SI{100}{\kilogram}), either provided in house or built by users. If a small sample environment is required (up to \SI{2}{\kilogram}), such as a single sample cell or lightweight capillary holder, 2 motorized translation stages equipped with stepper motors (Newport Spectra-Physics Ltd, Didcot, UK; UTS100PP) provide precise sample positioning with \SI{1}{\micro\metre} resolution over a \SI{100}{\milli\metre} range of motion horizontally and vertically.
The usual sample environments found on versatile SAXS beamlines are also available on I22. These can be controlled remotely, and trigger, or be triggered, by the data acquisition software. Both Linkam DSC and Capillary furnaces (Linkam Scientific, Surrey, UK) are available for use accessing temperatures in the range \SIrange{-180}{550}{\degreeCelsius}. Two commercial stopped flow apparatus (SM 400; Bio-Logic, Grenoble, France) specifically designed for synchrotron radiation SAXS are available. One is used for investigations of conformational changes in proteins \cite{Panine2006}, nucleic acids and macromolecules, while the second one is dedicated to material science studies \cite{Lund2013}. A stress controlled rheometer (Physica MCR 501; Anton Paar, Hertford, UK) is available with a range of geometries, suitable for a wide range of viscosities ranging from water to polymer melts \cite{Poulin2016}. It can be temperature controlled using air or water, providing an operating range between \SIrange[range-phrase = \ and\ ]{-20}{350}{\degreeCelsius}. A micromechanical tensile tester has been developed in collaboration with Queen Mary University of London \cite{Karunaratne2012b}. This can be used in tension, compression, and bending modes. A millisecond pressure-jump cell (\SIrange{1}{5000}{\bar}) for use in X-ray scattering experiments has been developed in collaboration with Imperial College London, and is now routinely available on I22 \cite{Brooks2010}. For higher pressure a Diamond Anvil cell is available for use with microfocus beam (Almax- easyLab, Diksmuide, Belgium; Boehler-Almax PlateDAC). This cell covers the pressure range \SIrange{1}{60}{\giga\pascal}.

\subsection{Detectors}

\begin{figure}
	\begin{center}
		\includegraphics[width=0.9\textwidth]{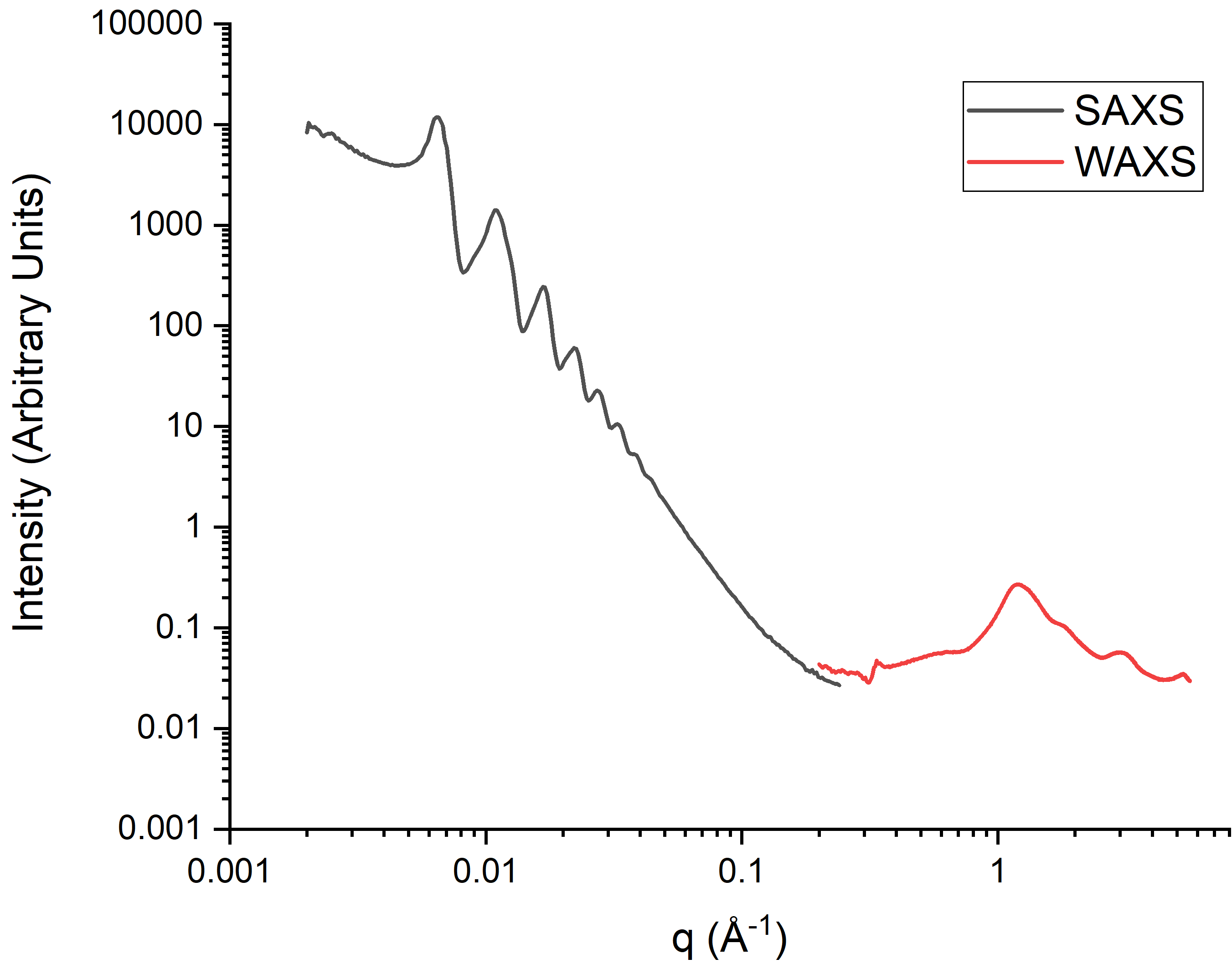}
		\caption{Overlapping q-scale of I22 detectors at 9m camera length, SAXS q-scale is \numrange[range-phrase = --]{0.002}{0.24} \si{\per\angstrom}, WAXS q-scale is \numrange[range-phrase = --]{0.2}{5.6} \si{\per\angstrom}. Data are from \SI{100}{\nano\metre} silica spheres collected as a powder between 2 Scotch Tape$^\text{TM}$ windows. NB: The WAXS data has been shifted upward slightly to better show the overlap region}\label{fg:qscale}
	\end{center}
\end{figure}

The beamline has a matched pair of Dectris (Dectris AG, Switzerland) Pilatus P3 Hybrid silicon pixel detectors \cite{Henrich2009}, see Table \ref{tb:detectors}. Of particular note is the in vacuum L shaped WAXS detector. This has the basic form of a Pilatus P3-2M with three modules in the lower right corner removed. The detector is built into the vacuum space of the SAXS camera nosecone affording partial 2D access to WAXS to complement full 2D SAXS in the appropriate configuration. With careful beamline set up, these detectors allow overlap of 1D SAXS and WAXS data over the full camera length capability of the beamline as highlighted by the data shown in figure \ref{fg:qscale}. It is possible to arrange the detectors such that no shadowing of the SAXS detector by the WAXS detector occurs.

% TS - To look into why this isn't centering
\begin{center}
    \begin{table}
        \caption{Detectors available on I22 and their main uses\\[1em]}
        \label{tb:detectors}
        \begin{tabular}{ll} % Alignment for each cell: l=left, c=center, r=right
            Detector & Use \\
            \hline
            Pilatus P3 2M & (Gi)SAXS Data Collection \\
             & \SI{172}{\micro\metre} pixel size \\
             & Total active area \SI{254}{\milli\metre} x \SI{289}{\milli\metre} \\
             & Frame rate \SI{250}{\hertz} with \SI{1}{\milli\second} read out \\
            Pilatus P3 2M-DLS-L &In vacuum (Gi)WAXS data collection \\
            & \SI{172}{\micro\metre} pixel size \\
            & Total active area \SI{254}{\milli\metre} x \SI{289}{\milli\metre} \\
            & Frame rate \SI{250}{\hertz} with \SI{1}{\milli\second} read out \\
            Vortex Single Element & Fluorescence detection for mapping experiments \\
            Silicon Pin Diode & Absorption correction data collection \\
             & \SI{1}{\milli\metre\squared} diode embedded in beamstop \\
        \end{tabular}
    \end{table}
\end{center}

\section{Software}

The beamline is controlled using an architecture based on the Experimental Physics and Industrial Control System (EPICS) \cite{Dalesio1994}. A graphical synoptic of the whole beamline is available and windows for specific components can be accessed to set and monitor process variables. Users interact with the beamline through the generic data acquisition client \cite{Rees2010,Gibbons2011} (GDA) a software-based data acquisition system that sits on top of the EPICS layer. For simple data collection, frames $\times$ collection length, there are a range of graphical user interfaces available and for more complex experiments the full Jython scripting capability of GDA can be employed.

After the acquisition of frame data, by default, GDA triggers any requested analyses that are available through the Data Analysis WorkbeNch (DAWN) software package \cite{Basham2015,Filik2017}. Most commonly, this analysis entails the reduction of frame data to a one-dimensional data set of intensity, \textit{I}, plotted against the scattering wave vector, \textit{q}, following a standardised schema \cite{Pauw2017a}. It is also possible to analyse particular Debye-Scherrer rings, producing \textit{I} versus azimuthal angle, \textit{$\chi$} plots, the calculation of orientation factors and degrees of orientation \cite{Cinader1998, Hermans1944} as well as crystallite thickness information from frame data \cite{Fratzl1996}. These scattering specific analyses are presented alongside a number of other common analytical methodologies, such as function fitting, any number of which can be automatically triggered following data acquisition.

\section{Science examples}

To illustrate the versatility and data quality of the beamline we represent below a selection of science highlights from the 10 years of operation of I22. These show the diversity of systems that can be studied on a flexibly designed SAXS/WAXS beamline as well as illustrate the range of sample environments available to users.

\subsection{Solution Scattering}\cite{OSullivan2011}

\noindent Large $\pi$-conjugated macrocycles are of interest for artificial photosynthesis applications because of their resemblance to natural light-harvesting complexes. Synthesis of these molecules is complicated by the need to template, or support, the structure during synthesis. The synthesis of larger target molecules requires commensurately larger scaffold molecules. Eventually synthesis of the scaffold molecule can become as challenging as the target molecule itself. Using the concept of a ``Vernier complex'', a combination of two molecular entities with an incommensurate number of complementary bindings sites, researchers from Oxford and Nottingham demonstrated that it is possible to build a porphyrin macrocycle in high yield, in a facile manner from easily synthesised building blocks. The target molecule was composed of 12 porphyrin units and synthesis was achieved from a combination of a linear building block of 4 porphyrin units, along with a circular hexapyridyl template. Two templates and three linear building blocks spontaneously form into a ``figure-of-eight'' Vernier complex. After coupling of the ends of the linear building blocks, the template can be removed to reveal the cyclic 12-unit porphyrin molecule. Solution scattering data collected at I22 confirmed the structures of both the ``figure-of-eight'', and the free cyclic molecule. At the time, the \SI{4.7}{\nano\metre} diameter 12-porphyrin nanoring was among the largest $\pi$-conjugated macrocycles synthesised. The group have continued to synthesise further related systems of increasing size and complexity. \cite{Sprafke2011, Cremers2018, Bols2018a}

\subsection{Colloid Science}\cite{Tang2014}

\noindent Cells are extremely complex systems and for life to form there are high demands for a chemically rich, highly structured, environment in the centre of the cell. Within the biological community there are, primarily, two competing theories that predict the formation of protocells, widely regarded as precursors to the cells we observe today and, therefore, a pre-requisite to early life on Earth. 

One thesis is based on fatty acids in water spontaneously forming a bilayer membrane in water which, in turn, subsequently self-assemble into fatty acid vesicles due to the partition function between the hydrophobic and hydrophilic elements of the molecule and water\cite{Deamer1982, Monnard2002}. The other is derived from the formation of chemically rich liquid microdroplets forming from omnipresent organic chemicals and natural polymers driving spontaneous phase separation generating a rich mixture of coacervates in solution\cite{Oparin1953}. 

Combining these two theses, researchers from the University of Bristol Centre for Protolife Research and the Chemistry Department of Imperial College London have examined a hybrid protocell model based on the spontaneous self-assembly of a fatty acid membrane on coacervate microdroplets.

Experiments conducted at I22 demonstrated that fatty acids assembled at the surface of the coacervate droplets forming highly organised multilayers. No multilamellar structures were seen in SAXS patterns at low fatty acid concentrations. Despite the low contrast between the two systems, the use of photon counting detectors on I22 meant that it was possible for Tang \textit{et al.} to determine the size and structure of the microcompartments, with some key features of interest being observed between \SIrange[range-phrase = \ and\   ]{4}{6}{\nano\metre} in size.

\subsection{Environmental Science}\cite{Bots2014}

\noindent There is a significant legacy of radioactive waste from nuclear power generation and military activities. Currently the favoured route for disposal of higher activity wastes is burial, deep underground, in a geological disposal facility (GDF) \cite{DEFRA2008} to contain the radionuclides until they can decay into stable forms. Almost all designs for GDFs involve cement and steel in their construction and backfill material. Once the facility is closed, and becomes saturated with groundwater, interaction between water and the cementitious material will generate a hyperalkaline leachate. Chemically reducing conditions are also expected to prevail in the facility, since any available oxygen will be locked up by corrosion of the steel components. The most abundant radionuclide expected in such a GDF will be uranium which under the conditions found in a GDF (high pH, slightly reducing) is most stable as the U(VI) ion. U(VI) adsorbs strongly to many solid surfaces and this is expected to limit the mobility of the radionuclide and help keep it within the bounds of the GDF. A team of researchers from Manchester and Leeds found that under the expected conditions of a GDF, uranium can form stable colloidal nanoparticles that might be expected to have much higher mobility and potentially escape from the GDF. Time-resolved SAXS data collected at I22 showed that these nanoparticles (1-2 nm in diameter) form within a few minutes, under conditions relevant to a GDF, and coalesce into colloidally stable aggregates within a few hours. The colloidal material remains stable for over 2.5 years. These results show that GDF design will need to account for the potential of uranium to form such nanoparticles and put in place measures to mitigate their migration from the GDF.

\subsection{Inorganic Chemistry}\cite{Stawski2016}

\noindent Calcium sulfate is a well-known and commonplace mineral, with large deposits of both the gypsum (CaSO$_4$·2H$_2$O) and anhydrite (CaSO$_4$) crystalline phases found naturally on Earth \cite{Warren2006}. A third phase, bassanite (CaSO$_4$·0.5H$_2$O), while of very limited abundance in natural settings, has an important role in the construction industry, with 100 billion tones of it produced annually as Plaster of Paris. Along with gypsum, bassanite has also recently been found as a natural deposit on Mars. \cite{Gendrin2005, Wray2010}
 
Despite its prevalence, the way in which gypsum forms from ions in solution remains a mystery. Gaining a better understanding of the fundamental nucleation and growth of the mineral is key if scientists want to control its formation. Ultimately this could lead to more energy efficient routes of making bassanite, a process which is currently very energy intensive.
 
Previous studies have used various Transmission Electron Microscopy approaches to follow gypsum’s growth \cite{Wang2012a, Driessche2012}, however this technique requires samples to be frozen or dried and studied under a vacuum, where each process may introduce unintended artefacts. Mindful of this problem, a pan-European team of researchers from institutes in the UK, Germany, Spain, Belgium, France, and Denmark, used I22 to develop a method that was truly \textit{in situ}, and followed the nucleation and crystallisation in solution and in real time.

Stawski \textit{et. al.} found that the crystallisation proceeds via a four step mechanism, which begins with the formation of ``nano-bricks'' - amorphous CaSO$_4$ cores. These are well dispersed throughout the water matrix. In the second step the bricks move slowly closer to each other, beginning to form unordered domains and creating a disordered mosaic. Only in step three do the nanobricks start to self-assemble into a wall like structure, with them crystallising into gypsum in the fourth and final step.

\subsection{Liquid Crystals}\cite{Zeng2016}

\noindent Research into the formation and structure of liquid crystal systems has delivered a myriad of technological and societal advances. Despite the maturity of this field new exotic compositions and morphologies are continually being discovered facilitating exciting possibilities in display technologies, \cite{Xiang2015, Wang2016} passive thermal stabilisation \cite{Ptasinski2014} and adaptive optics. \cite{Arines2009, Harada2018} Concomitantly, such studies have also extended our understanding of the mechanics within cellular membranes \cite{Lee2017} and the formation of certain biological materials, such as silk. \cite{Walker2015}

Advancing this domain further, researchers from the University of Sheffield conducted a study revealing a novel liquid crystal structure where bolaamphiphilic molecules were found to self-assemble into a $Pn\bar{3}m$ double diamond cubic liquid-crystalline phase. This work was the first unambiguous demonstration of a thermotropic double diamond phase opening the door to a family of new three-dimensional liquid crystal structured based around the design principle of skeletal networks consisting of axial mesogen bundles. % Double check space group

Measurements conducted on I22 were instrumental in the identification of this new structural family via their morphological characterisation, in both the SAXS and WAXS regimes, complementing previously obtained polarimetry and differential scanning calorimetry measurements. The combination of the results presented in these datasets together allowed Zeng \textit{et al.} to conclusively assign this phase, despite the unexpected structure uncovered.

\subsection{Biomaterials}\cite{Inamdar2017}

\noindent Articular cartilage is a natural biomaterial whose structure at the micro- and nano-scale is critical for healthy joint function and where degeneration is associated with widespread disorders such as osteoarthritis. At the nanoscale, cartilage mechanical functionality is dependent on the collagen fibrils and hydrated proteoglycans that form the extracellular matrix. \cite{Setton1999, Korhonen2002} The observation of the dynamic response of these ultrastructural building blocks at the nanoscale has, however, proven elusive.

Inamdar \textit{et al.} were able, for the first time, to observe the dynamics via their utilisation of highly time-resolved, microfocussed, tensile stress and strain SAXS measurements performed on I22. Using a micro-mechanical tester \cite{Karunaratne2012b} it was possible to observe the time-resolved changes in collagen fibrillar D-periodicity during the compression of both bovine and human cartilage explants. 

Results obtained demonstrated the existence of a pre strain in the collagen fibril, estimated from the D-period at approximately 1-2\%. It is thought that this is due to osmotic swelling pressure induced by the negatively charged proteoglycans.

Uniquely, the data obtained shows a rapid reduction and recovery of this pre-strain which occurs during stress relaxation, approximately 60 seconds after the onset of peak load. Furthermore, the results highlight that this reduction in pre-strain is linked to disordering in the intrafibrillar molecular packing, alongside changes in the axial overlapping of tropocollagen molecules within the fibril. 

Inamdar \textit{et al.} observed that tissue degradation in the form of selective proteoglycan removal disrupts both the collagen fibril pre-strain and the transient response during stress relaxation. This study bridges a fundamental gap in the knowledge describing time-dependent changes in collagen pre-strain and molecular organisation that occur during physiological loading of articular cartilage. 

This previously unknown transient response is likely to transform our understanding of the role of collagen fibril nano-mechanics in the biomechanics of cartilage and other hydrated soft tissues.

\subsection{Nanostructures}\cite{Brady2017}

There are a wide range of emerging technological applications \cite{Rothemund2006, Iinuma2014, Ke2012, Tian2015, Jones2015} that require materials with a precisely defined 3 dimensional nanostructure. To achieve this the structures must be built from the ``bottom-up'' and, since the molecular building blocks are too small to manipulate directly, they must be encouraged to self-assemble. 

Many soft matter systems feature self-assembly but directing this towards desired structures is problematic, particularly if the desired structure has significant porosity. One class of materials that have the potential to deliver bespoke porous 3D nanostructures are those made from self-assembled DNA, dubbed ``nanostars''. Despite their initial promise it has been found that the DNA nanostructures suffer from a lack of flexibility that hinders their ability to form contiguous crystalline networks. A team of researchers from the University of Cambridge and Imperial College London proposed a modification to the system that would relax the strict geometrical constraints that prevent the DNA nanostructures from realising their potential. 

By adding strands of cholesterol to the ends of the DNA strands they were able to create a range of highly porous ``C-stars''. It is the amphiphilic nature, and variable binding valency of the cholesterol sections that give the C-stars the required flexibility to form the intricate 3-d networks. Both bulk powder SAXS and microfocus single crystal SAXS at I22 were used by the team to characterise the structures. They continue to explore the diverse structures that this technology has enabled.

\section{Summary}

I22 was designed as a multipurpose, undulator driven, combined SAXS-WAXS facility to serve the needs of a broad UK community of SAXS/WAXS users. The beamline was, therefore, designed as a very versatile platform capable of accommodating a wide variety of biological and soft matter studies. The range of science covered has grown during its operation to include users supported by all of the major UK Science Research Councils as well as European funding bodies and beyond. Being Diamond’s primary multipurpose SAXS beamline, it has been required to carry out studies ranging from studying precursors to life on earth \cite{Tang2014}, through self-assembly of polymer thin films \cite{Pearson2014}, to microfocus SAXS mapping studies on bone \cite{Karunaratne2013} and cornea for a range of medical studies (e.g. \cite{Hayes2013}). The selected examples illustrate the potential of the I22 beamline as a high flux, high resolution, X-ray scattering beamline for \textit{in situ} and time resolved investigations.

     %-------------------------------------------- -----------------------------
     % The back matter of the paper - acknowledgements and references
     %-------------------------------------------------------------------------

     % Acknowledgements come after the appendices

\ack{Acknowledgements} % Formatting will be done by the journal

We would like to thank the large number of Diamond Light Source personnel, both past and present, that have been involved in the detailed design, build, and commissioning of beamline I22 and its components. In particular we would like to thank Alan Bone, Tom Cobb, Alan Day, Alan Grant, Andy Gundry, Ian Johnson, Simon Lay, Pete Leicester, Emily Longhi, Ronaldo Mercado, Brian Nutter, Mark Popkiss, Tobias Richter, Chris Roper, Irakli Sikharulidze, Andy Smith, Richard Staunton-Lambert, Andrew Thompson, Kevin Wilkinson, Joe Williams, and Faijin Yuan.

\referencelist[BeamlinePaper.bib]
   
\end{document}

% DO NOT DELETE THIS LINE

     %-------------------------------------------------------------------------
     % The introductory (header) part of the paper
     %-------------------------------------------------------------------------

     % The title of the paper. Use \shorttitle to indicate an abbreviated title
     % for use in running heads (you will need to uncomment it).

\title{Supplementary Information - I22: SAXS/WAXS beamline at Diamond Light Source - an overview of 10 years operation}
%\shorttitle{Short Title}

     % Authors' names and addresses. Use \cauthor for the main (contact) author.
     % Use \author for all other authors. Use \aff for authors' affiliations.
     % Use lower-case letters in square brackets to link authors to their
     % affiliations; if there is only one affiliation address, remove the [a].

\author[a]{A. J.}{Smith}
\author[a]{S. G.}{Alcock}
\author[a]{L. S.}{Davidson}
\author[a]{J. H.}{Emmins}
\author[d]{J. C.}{Hiller Bardsley}
\author[a]{P.}{Holloway}
\author[c]{M.}{Malfois}
\author[a]{A. R.}{Marshall}
\author[a]{C. L.}{Pizzey}
\author[b]{S. E.}{Rogers}
\author[a]{O.}{Shebanova}
\author[a]{T.}{Snow}
\author[a]{J. P.}{Sutter}
\author[a]{E. P.}{Williams}
\cauthor[a]{N. J.}{Terrill}{nick.terrill@diamond.ac.uk}{address if different from\aff}

\aff[a]{Diamond Light Source Ltd, Diamond House, Harwell Science and Innovation Campus, Didcot, Oxfordshire, OX11 0DE, \country{United Kingdom}}
\aff[b]{ISIS Neutron and Muon Source, Science and Technology Facilities Council, Rutherford Appleton Laboratory, Didcot, Oxfordshire, OX11 0QX, \country{United Kingdom}}
\aff[c]{ALBA Synchrotron, Carrer de la Llum 2-26, 08290 Cerdanyola del Vall\`{e}s, Barcelona, \country{Spain}}
\aff[d]{King's College London, Guy's Campus, Great Maze Pond, London SE1 1UL \country{United Kingdom}}

     % Use \shortauthor to indicate an abbreviated author list for use in
     % running heads (you will need to uncomment it).

%\shortauthor{Soape, Author and Doe}

     % Use \vita if required to give biographical details (for authors of
     % invited review papers only). Uncomment it.

%\vita{Author's biography}

     % Keywords (required for Journal of Synchrotron Radiation only)
     % Use the \keyword macro for each word or phrase, e.g. 
     % \keyword{X-ray diffraction}\keyword{muscle}

%\keyword{keyword}

     % PDB and NDB reference codes for structures referenced in the article and
     % deposited with the Protein Data Bank and Nucleic Acids Database (Acta
     % Crystallographica Section D). Repeat for each separate structure e.g
     % \PDBref[dethiobiotin synthetase]{1byi} \NDBref[d(G$_4$CGC$_4$)]{ad0002}

%\PDBref[optional name]{refcode}
%\NDBref[optional name]{refcode}

\maketitle                        % DO NOT DELETE THIS LINE

     %-------------------------------------------------------------------------
     % The main body of the paper
     %-------------------------------------------------------------------------
     % Now enter the text of the document in multiple \section's, \subsection's
     % and \subsubsection's as required.

\section{Calibration EXAFS spectra and derivatives}
\beginsupplement

\begin{figure}
	\begin{center}
	    \includegraphics[width=0.9\textwidth]{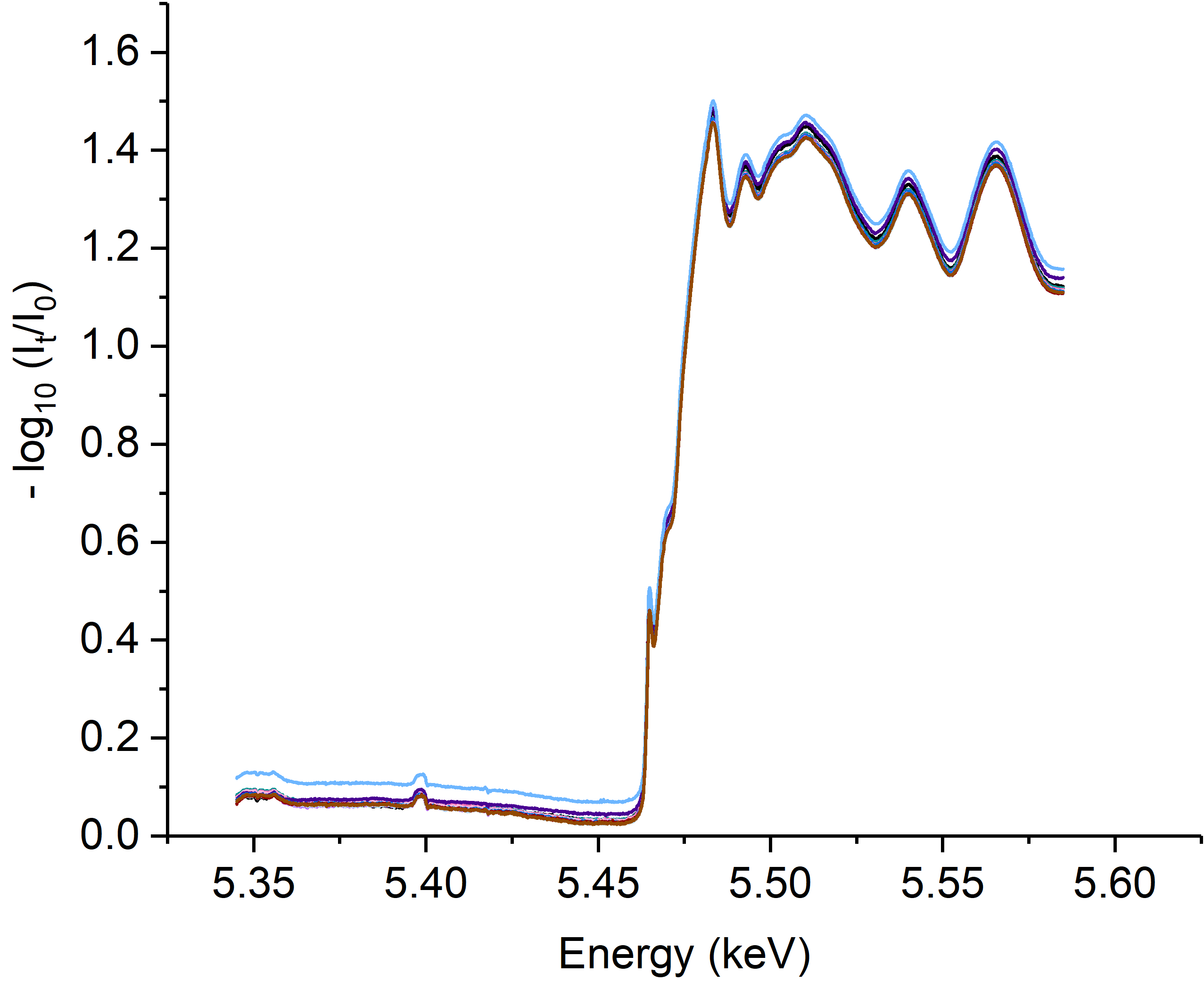}
	    \includegraphics[width=0.9\textwidth]{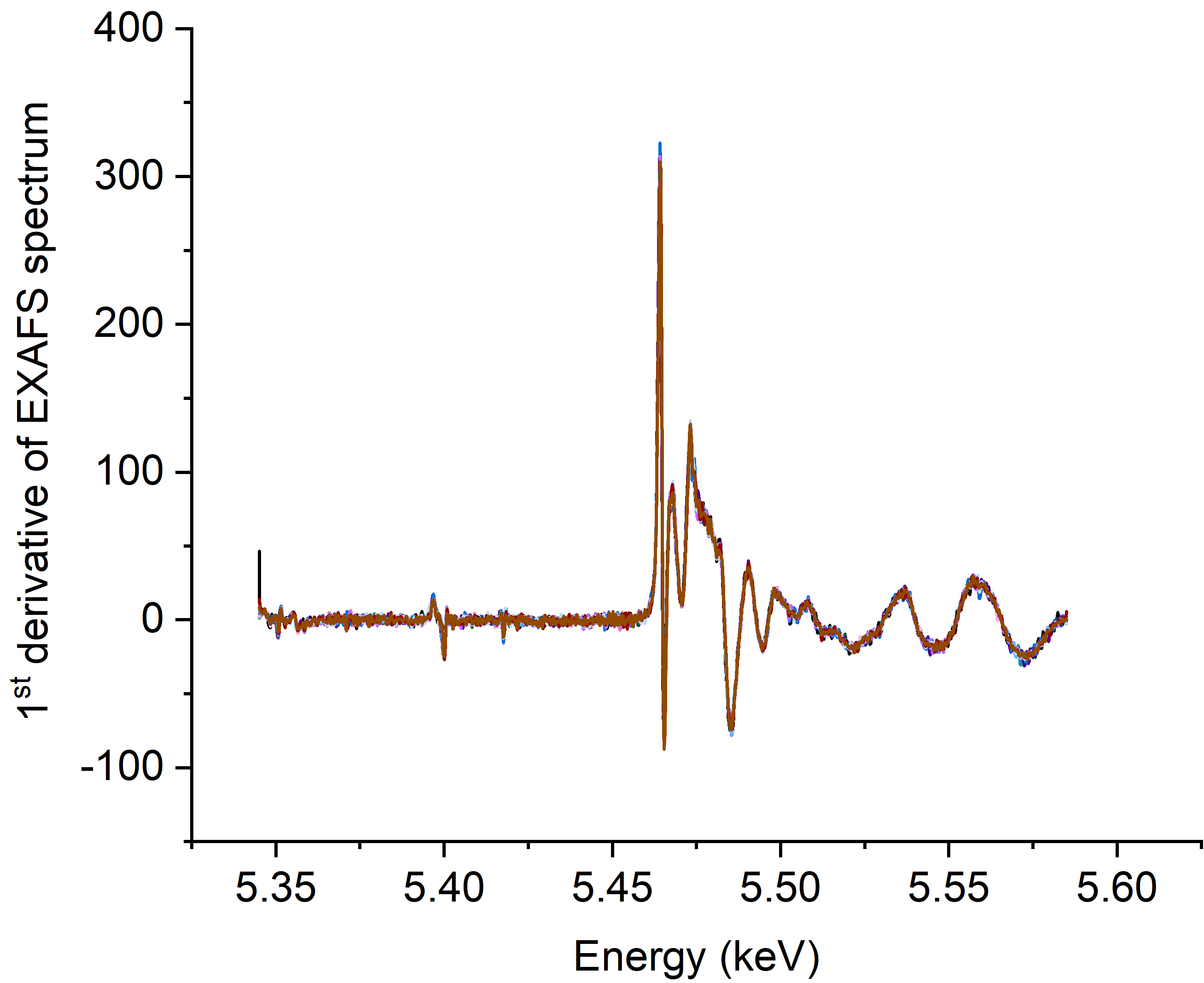}
		\caption{Vanadium K edge EXAFS @ 0.2 eV resolution}\label{fg:V_EXAFS}
	\end{center}
\end{figure}

\begin{figure}
	\begin{center}
	    \includegraphics[width=0.9\textwidth]{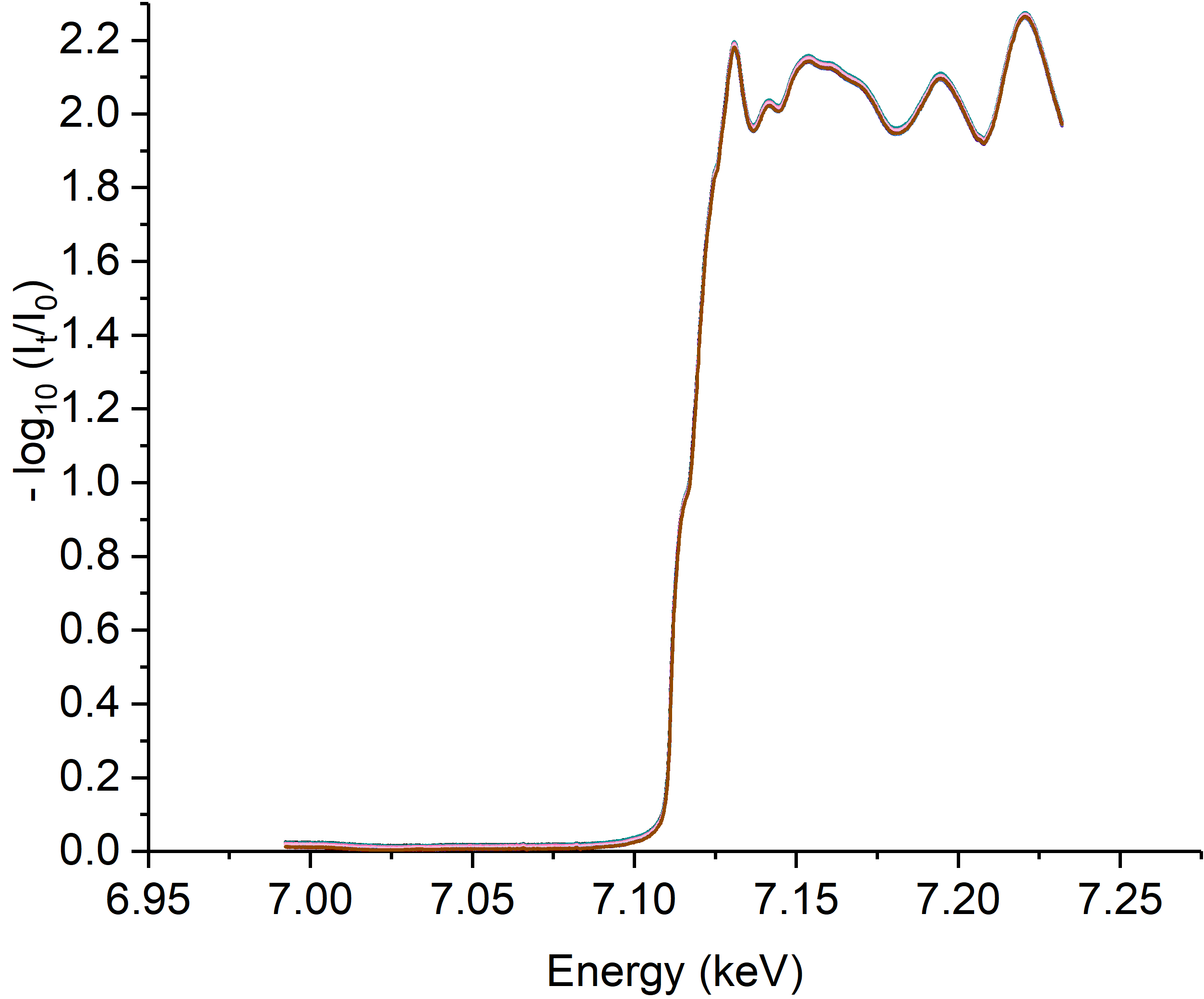}
	    \includegraphics[width=0.9\textwidth]{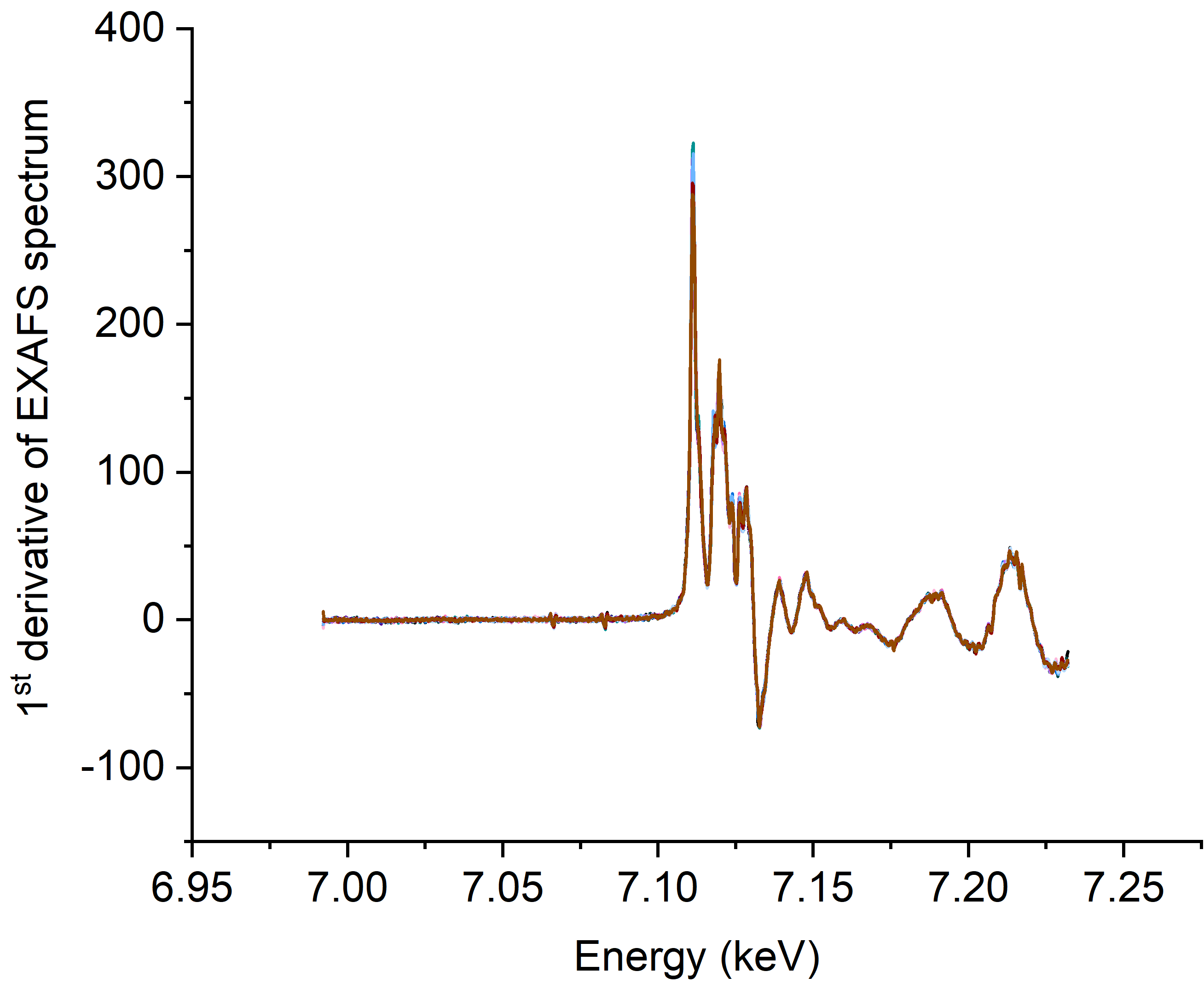}
		\caption{Iron K edge EXAFS @ 0.2 eV resolution}\label{fg:Fe_EXAFS}
	\end{center}
\end{figure}

\begin{figure}
	\begin{center}
	    \includegraphics[width=0.9\textwidth]{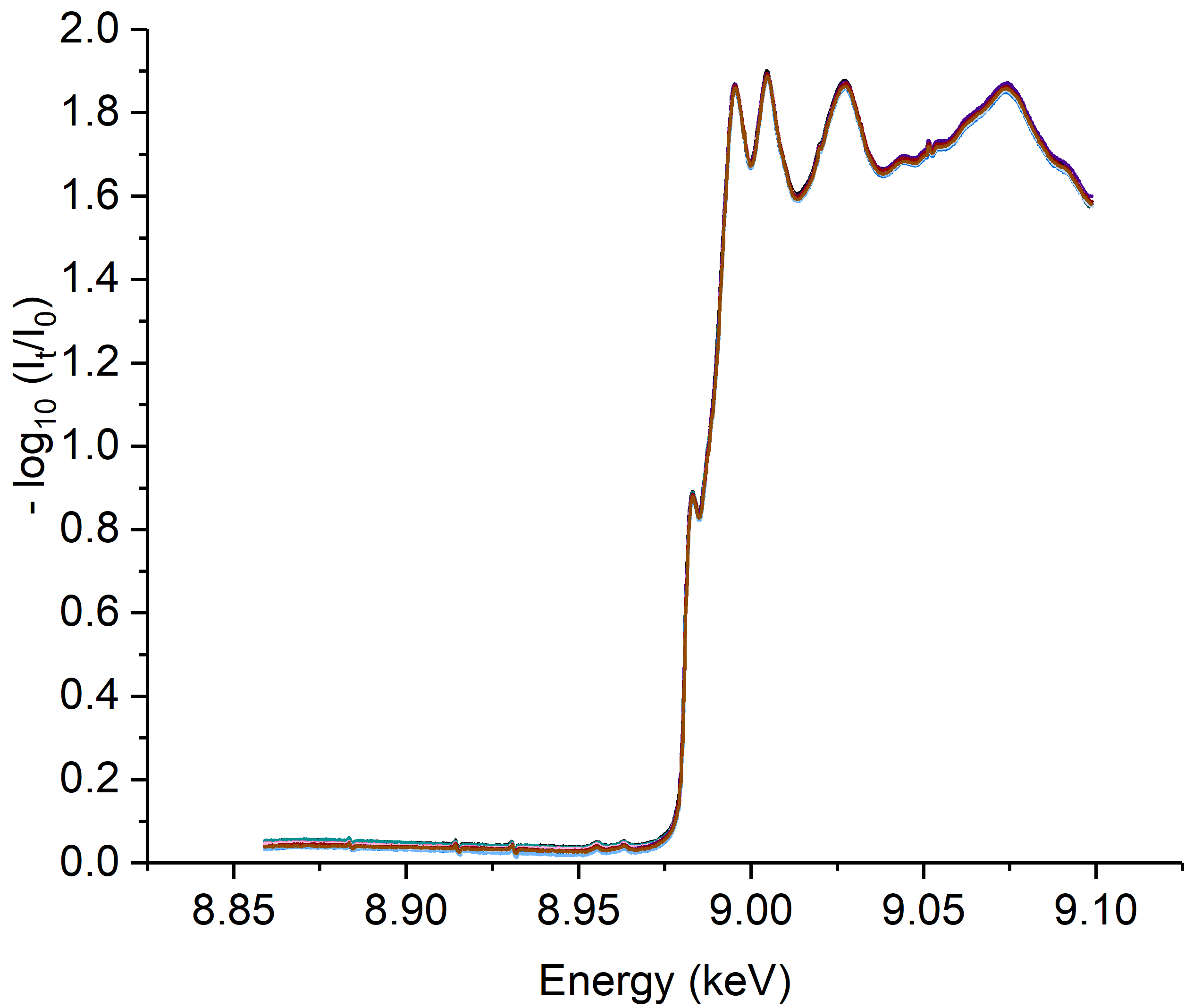}
	    \includegraphics[width=0.9\textwidth]{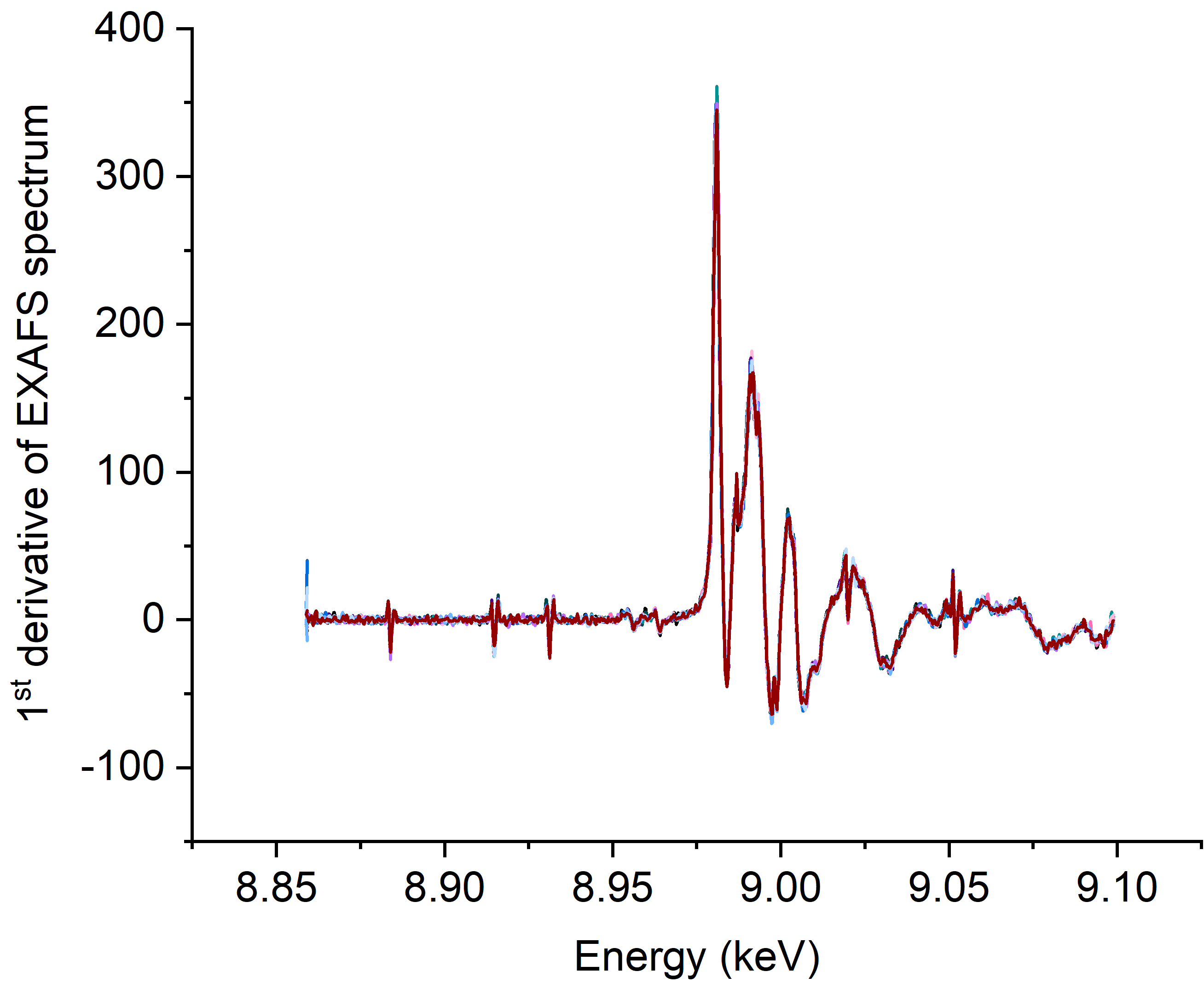}
		\caption{Copper K edge EXAFS @ 0.2 eV resolution}\label{fg:Cu_EXAFS}
	\end{center}
\end{figure}

\begin{figure}
	\begin{center}
	    \includegraphics[width=0.9\textwidth]{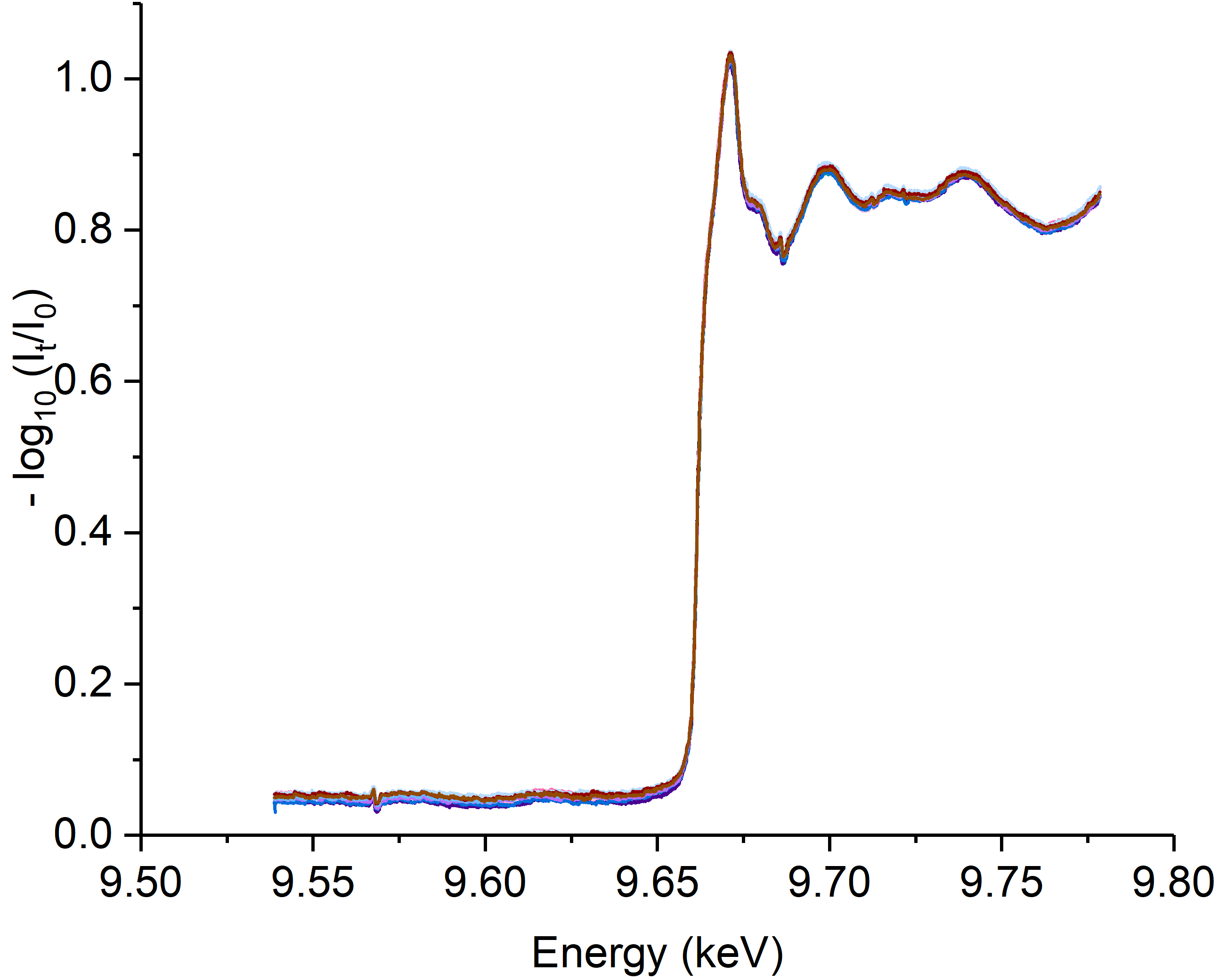}
	    \includegraphics[width=0.9\textwidth]{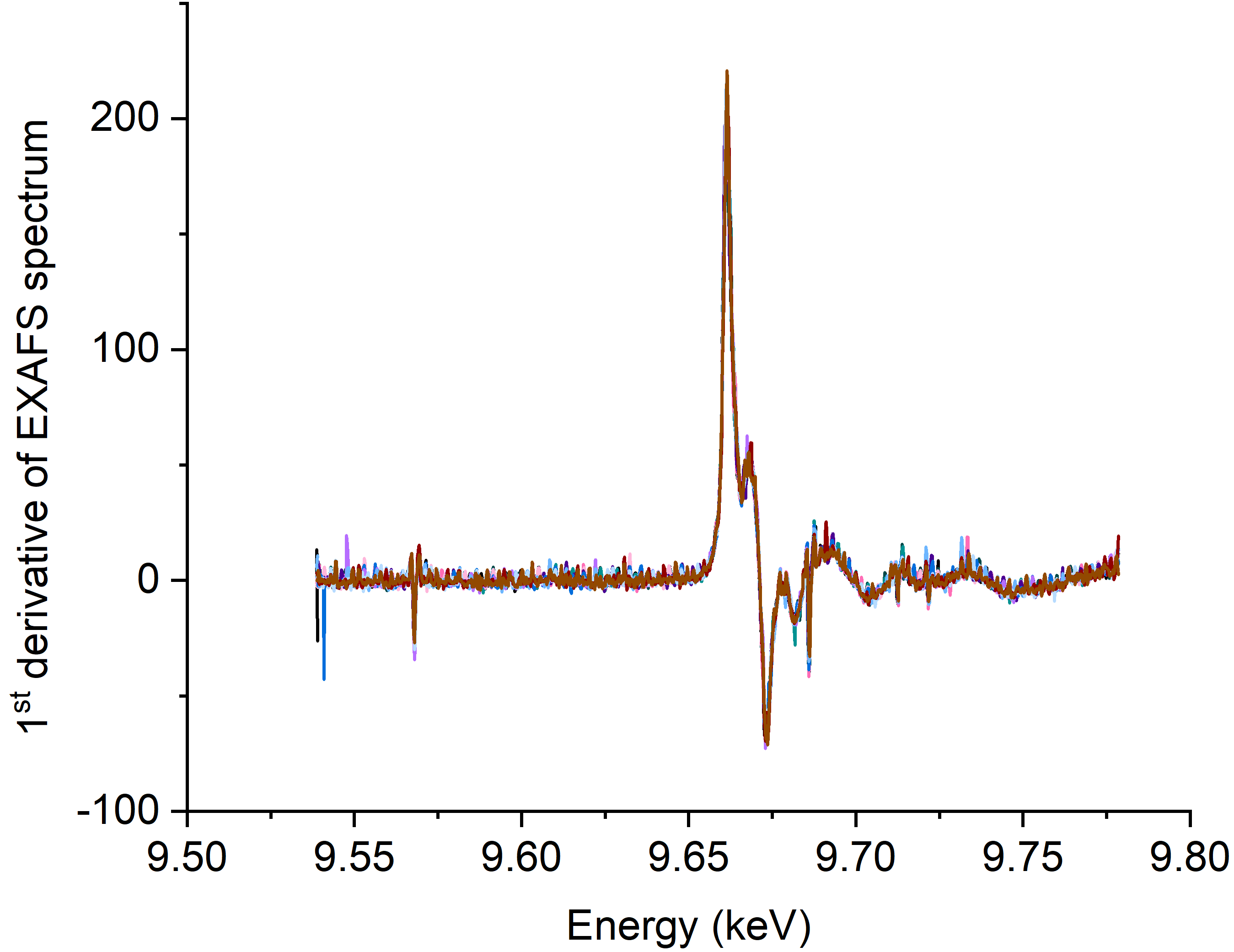}
		\caption{Zinc K edge EXAFS @ 0.2 eV resolution}\label{fg:Zn_EXAFS}
	\end{center}
\end{figure}

\begin{figure}
	\begin{center}
	    \includegraphics[width=0.9\textwidth]{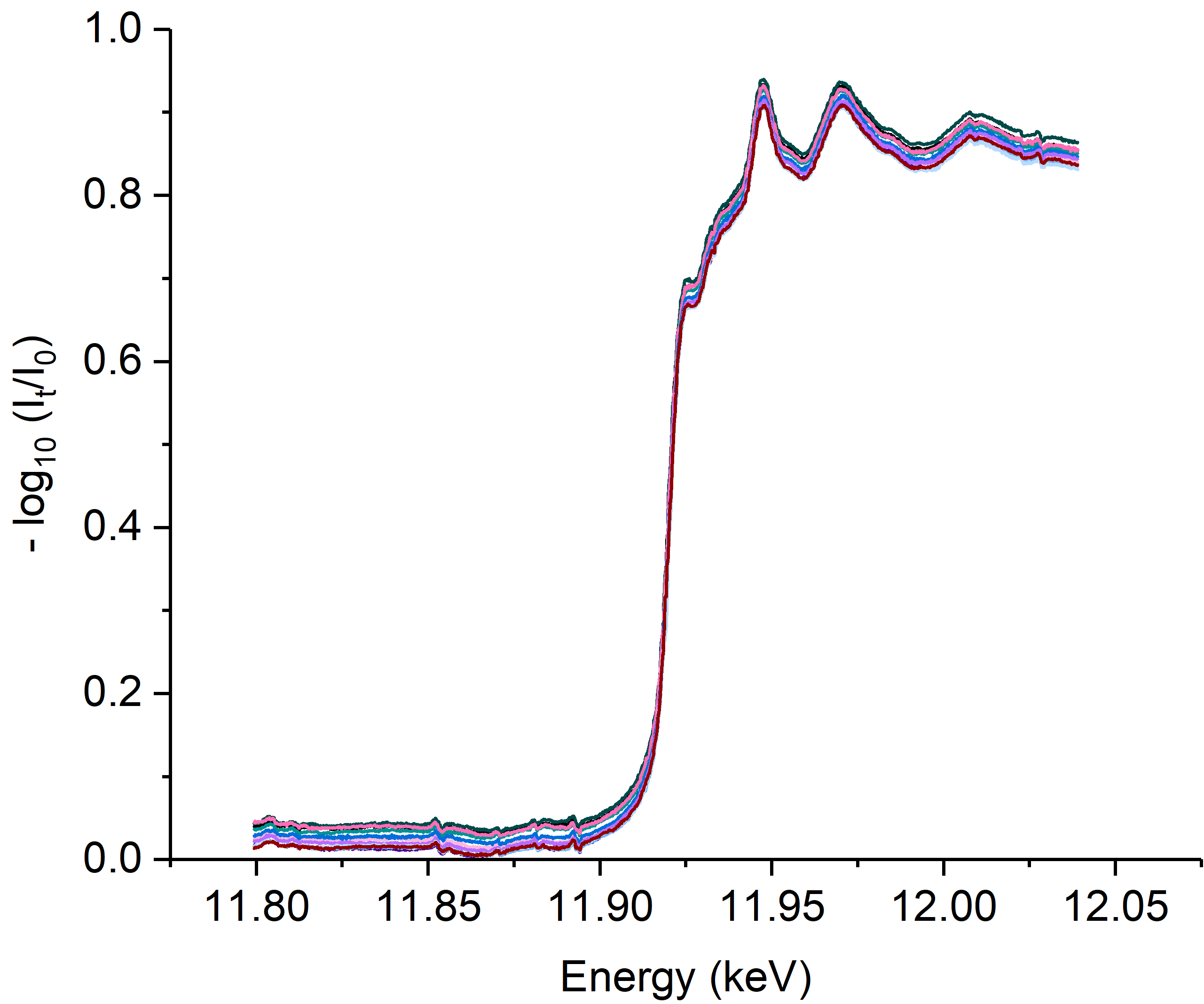}
	    \includegraphics[width=0.9\textwidth]{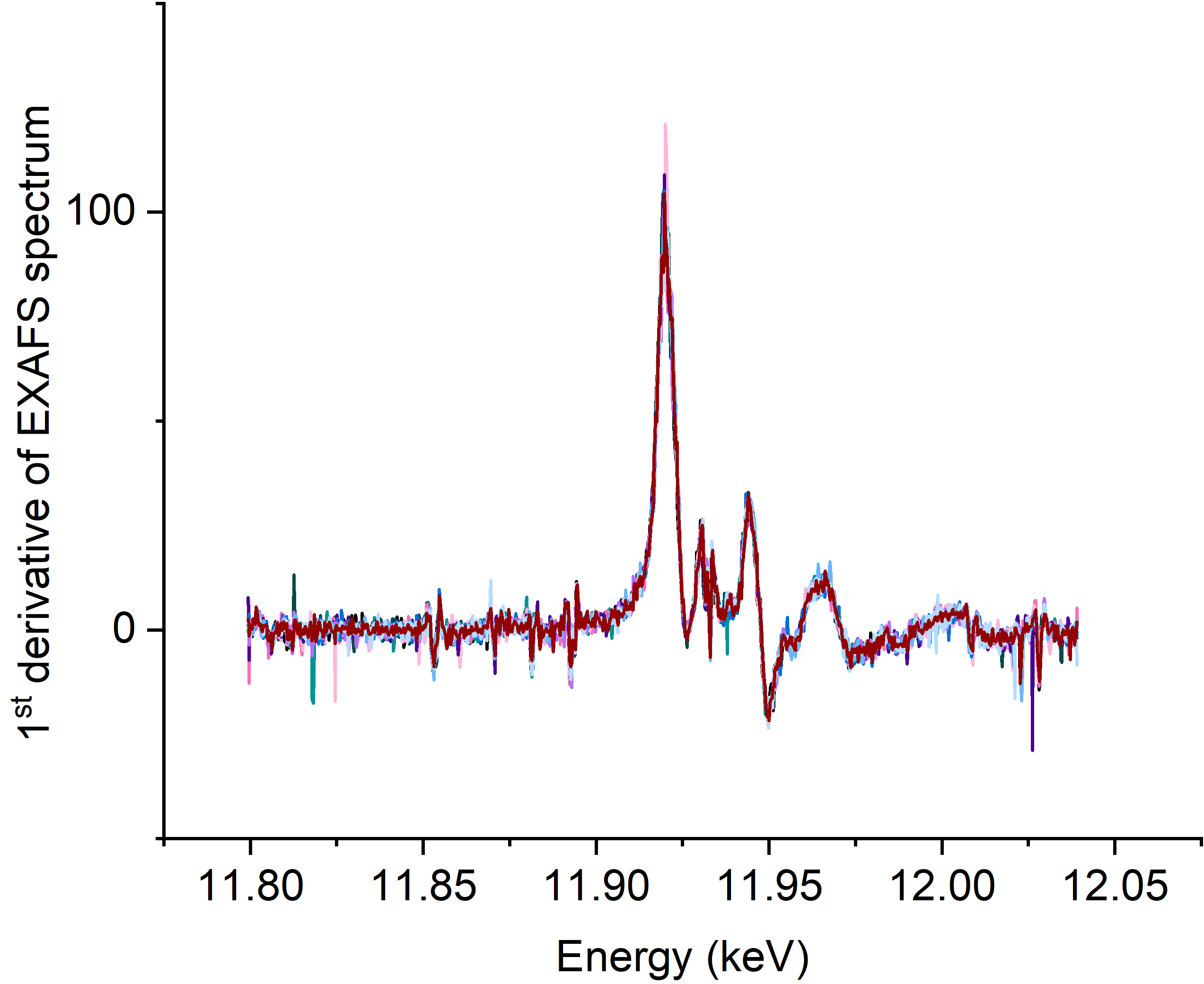}
		\caption{Gold LIII edge EXAFS @ 0.2 eV resolution}\label{fg:Au_EXAFS}
	\end{center}
\end{figure}

\begin{figure}
	\begin{center}
	    \includegraphics[width=0.9\textwidth]{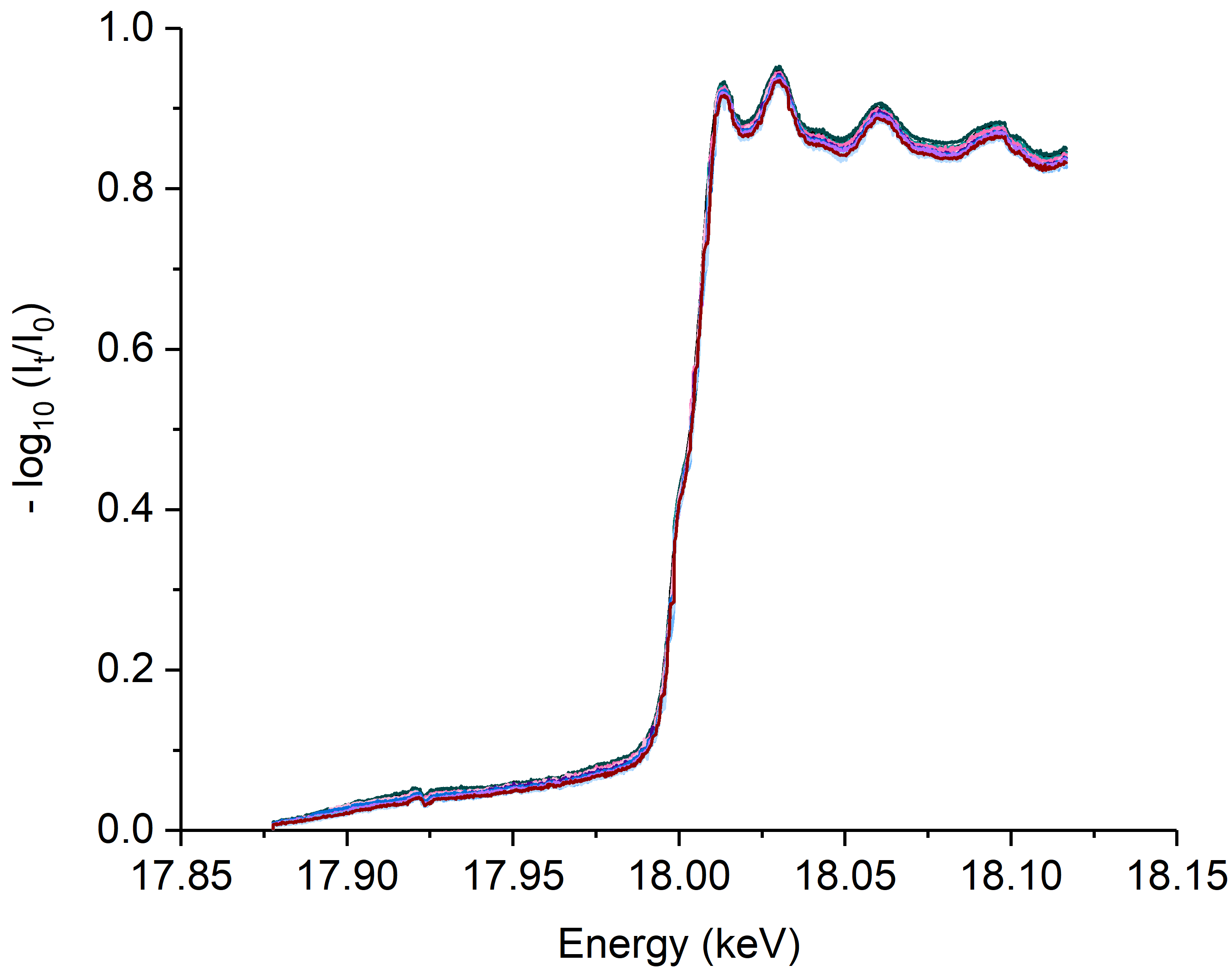}
	    \includegraphics[width=0.9\textwidth]{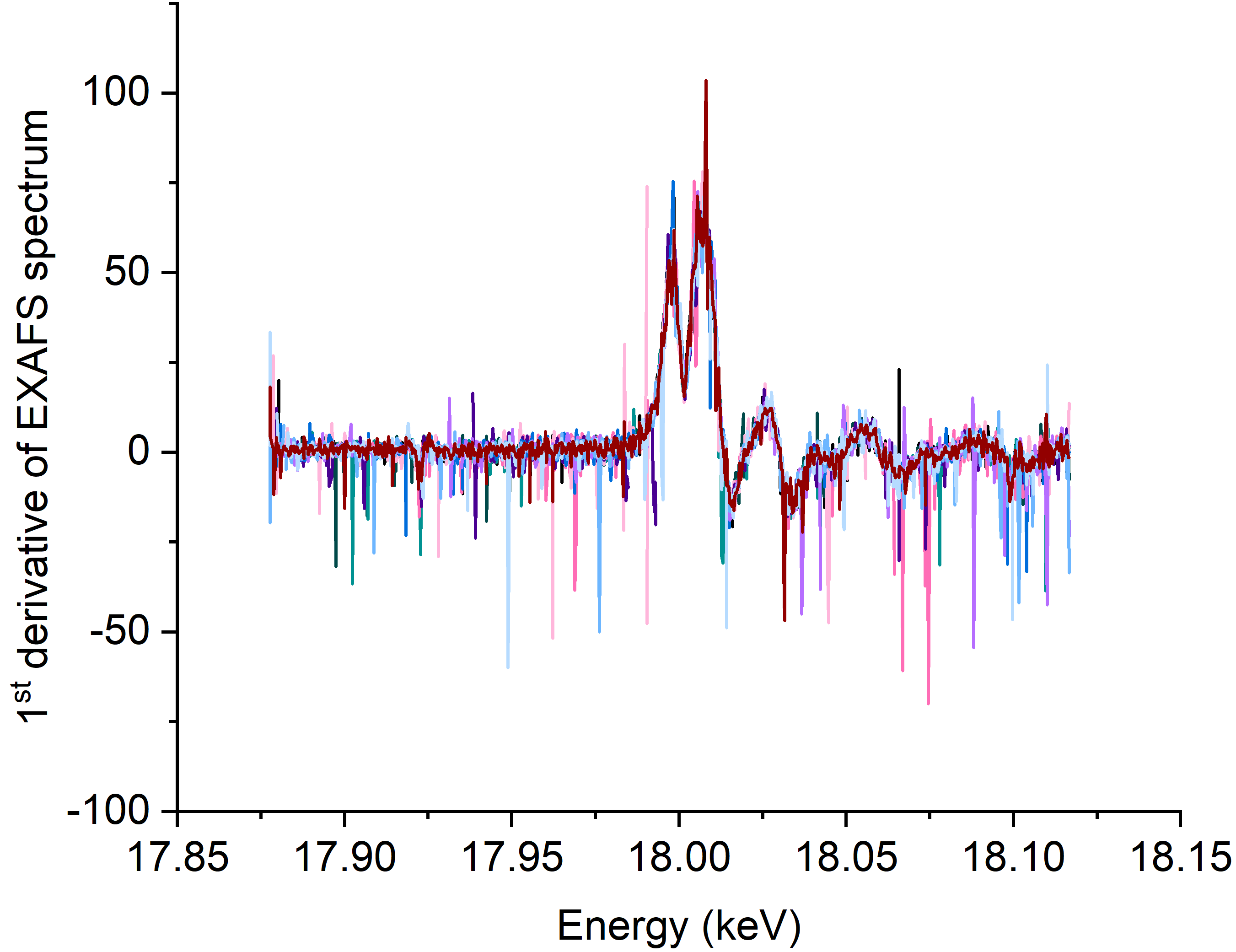}
		\caption{Zirconium K edge EXAFS @ 0.2 eV resolution}\label{fg:Zr_EXAFS}
	\end{center}
\end{figure}

\begin{figure}
	\begin{center}
	    \includegraphics[width=0.9\textwidth]{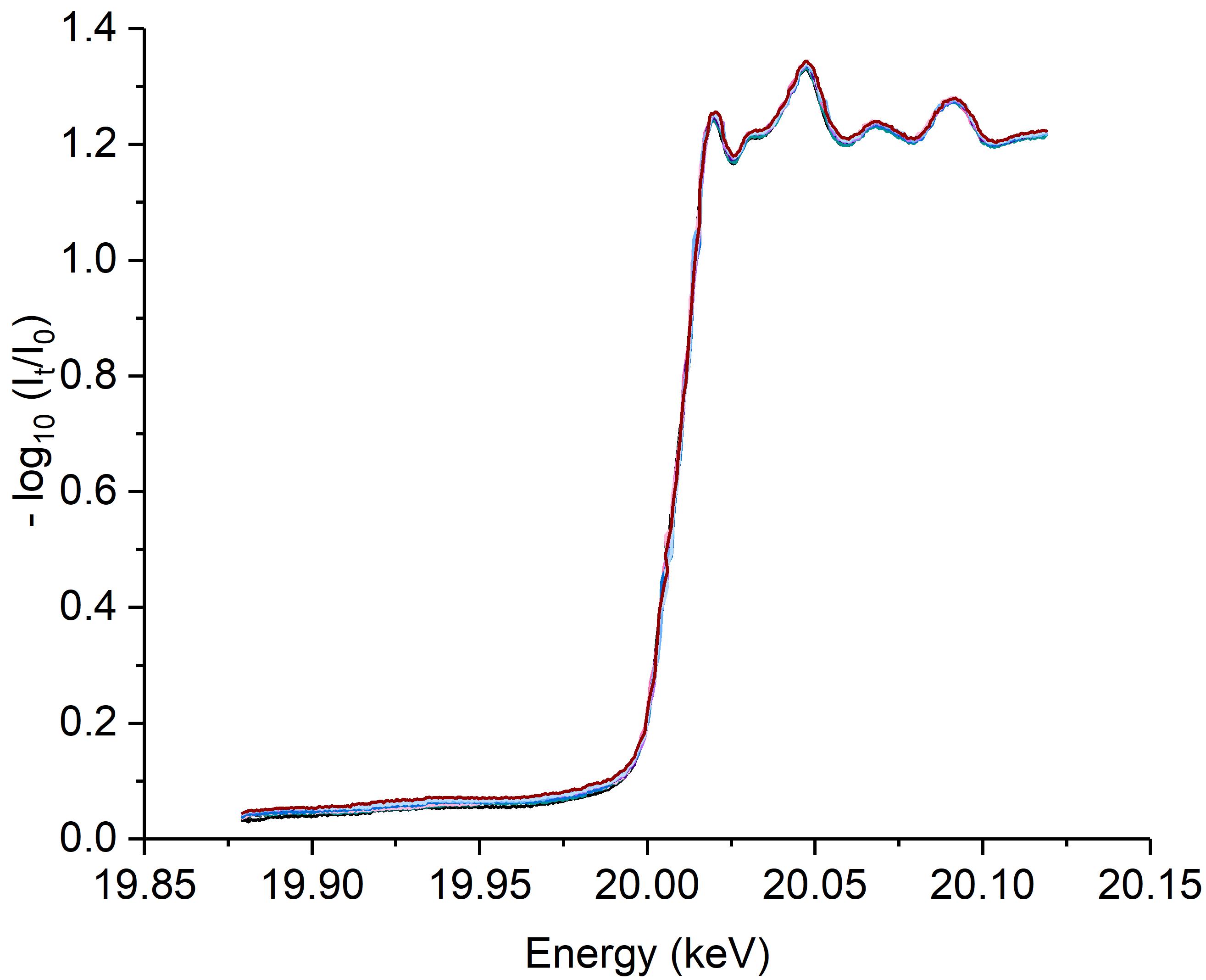}
	    \includegraphics[width=0.9\textwidth]{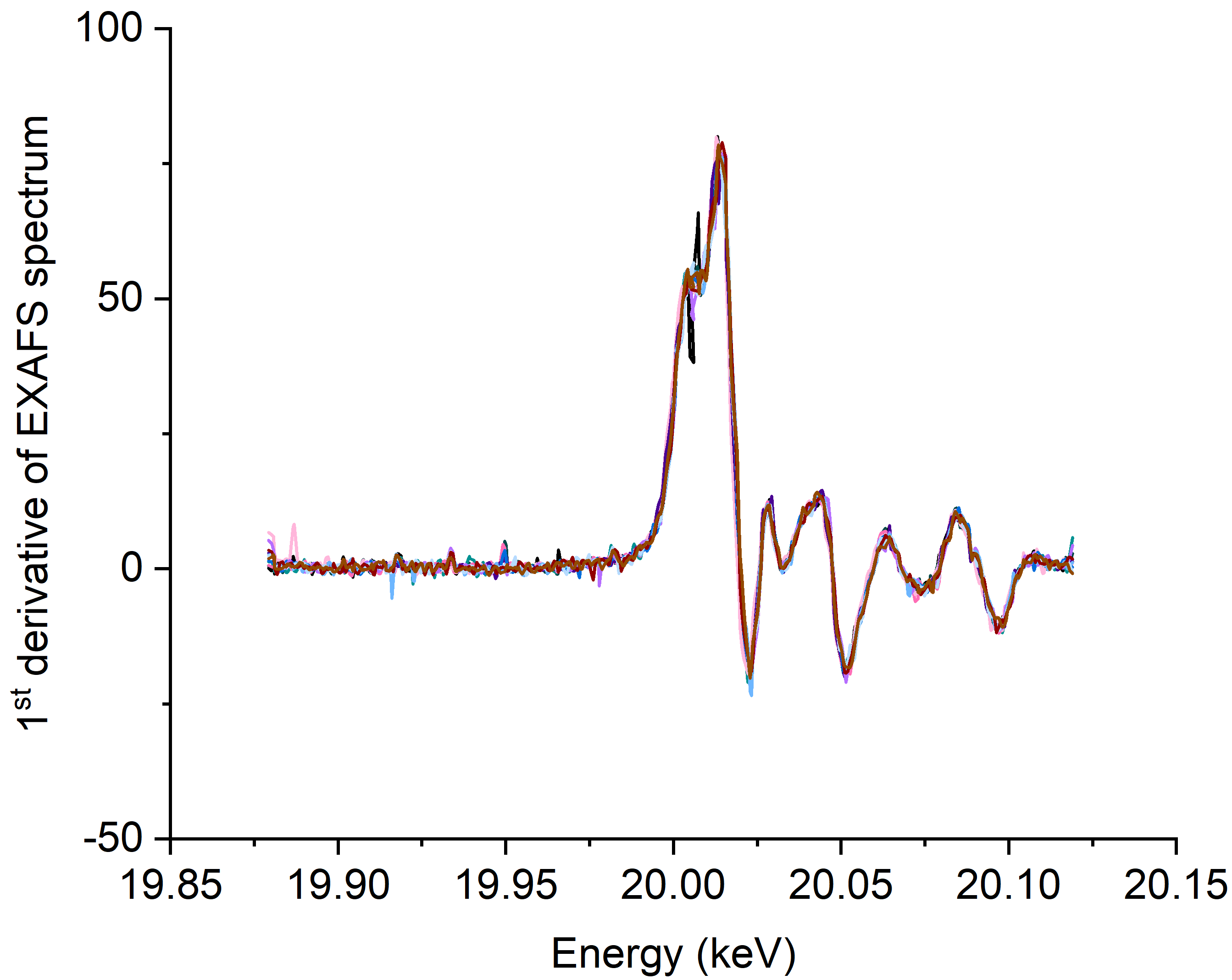}
		\caption{Molybdenum K edge EXAFS @ 0.5 eV resolution}\label{fg:Mo_EXAFS}
	\end{center}
\end{figure}